\documentclass[sigconf, 10pt, nonacm=true, balance=false]{acmart}

\usepackage{listings}
\usepackage{color}
\usepackage{float}
\renewcommand{\abstractname}{Summary}

\definecolor{codegreen}{rgb}{0,0.6,0}
\definecolor{codegray}{rgb}{0,0,0}
\definecolor{codepurple}{rgb}{0.58,0,0.82}
\definecolor{backcolour}{rgb}{0.95,0.95,0.92}
\usepackage{xspace}

\newcommand{\sys}{\mbox{\textsc{LEHAR}}\xspace}
 
\lstdefinestyle{mystyle}{
    backgroundcolor=\color{backcolour},   
    commentstyle=\color{codegreen},
    keywordstyle=\color{magenta},
    numberstyle=\tiny\color{codegray},
    stringstyle=\color{codepurple},
    basicstyle=\fontsize{8}{10}\ttfamily,
    breakatwhitespace=false,         
    breaklines=true,                 
    captionpos=b,                    
    keepspaces=true,                 
    numbers=left,                    
    numbersep=5pt,                  
    showspaces=false,                
    showstringspaces=false,
    showtabs=false,                  
    tabsize=2,
    frame=single
}
 
\def\BibTeX{{\rm B\kern-.05em{\sc i\kern-.025em b}\kern-.08emT\kern-.1667em\lower.7ex\hbox{E}\kern-.125emX}}

\begin{document}

%
\title{Enabling High-Accuracy Human Activity Recognition with Fine-Grained Indoor Localization}

%
\author{Arvind Seshan}
\affiliation{%
  \city{Pittsburgh}
  \state{PA, USA}
}

%
\renewcommand{\shortauthors}{Seshan}

%

\begin{abstract}
While computers play an increasingly important role in every aspect of our lives, their inability to understand what tasks users are physically performing makes a wide range of applications, including health monitoring and context-specific assistance, difficult or impossible. With Human Activity Recognition (HAR), applications could track if a patient took his pills and detect the behavioral changes associated with diseases such as Alzheimer’s. Current systems for HAR require diverse sensors (e.g., cameras, microphones, proximity sensors, and accelerometers) placed throughout the environment to provide detailed observations needed for high-accuracy HAR. The difficulty of instrumenting an environment with these sensors makes this approach impractical.  

This project considers whether recent advances in indoor localization (Wi-Fi Round Trip Time) enable high-accuracy HAR using only a smartphone. My design, called Location-Enhanced HAR (LEHAR), uses machine learning to combine acceleration, audio, and location data to detect common human activities. A LEHAR prototype, designed to recognize a dozen common activities conducted in a typical household, achieved an $F_1$-score of 0.965. In contrast, existing approaches, which use only acceleration or audio data, obtained $F_1$-scores of 0.660 and 0.865, respectively, on the same activities. In addition, the $F_1$-score of existing designs dropped significantly as more activities were added for recognition, while LEHAR was able to maintain high accuracy. The results show that using a combination of acceleration, audio, and Wi-Fi Round Trip Time localization can enable a highly accurate and easily deployable HAR system.

\end{abstract}

%
\keywords{Human Activity Recognition (HAR), Wi-Fi RTT (Round Trip Time), neural network, Root-mean-square Energy, Chroma Short-Time Fourier Transform, Spectral Centroid, Spectral Rolloff, Spectral Bandwidth, Zero Crossing Rate, Mel Frequency Cepstral Coefficients, confusion matrix, EWMA
}

%
\maketitle

\section{Introduction}
\label{sec:intro}


Computing devices today have little awareness of their surroundings. This shortcoming is slowly being addressed by the addition of many different sensor types to mobile devices such as smartphones. The added sensors have enabled a variety of popular applications ranging from GPS-based navigation to step-counting. However, these sensors and enabled applications are just a stepping stone towards having computing devices understand what physical activities users are performing and providing them useful guidance in these tasks. 


There are many potential applications of Human Activity Recognition or HAR.  For example, HAR can be used in a device that could determine that a user is cooking and could prompt you with recipes and even step-by-step guidance in the preparation. HAR could also play an important role in health care. An HAR system could assist senior citizens by analyzing patterns in activities over long periods of time and look for discontinuities such as if he or she forgot to take their medicine one day. This could help seniors achieve greater independence in their daily lives and use changes in behavior to identify the onset of potential diseases.

Unfortunately, these applications have been hampered by the fact that accurately identifying what activity a user is performing has proven difficult for a variety of reasons. At its core, HAR involves the classification of human activities through the analysis of sensor data. This, in turn, creates two sub-problems: 1) the collection of sensor data, and 2) the classification into activities. 

The sensor data necessary for HAR can come from a variety of sources and existing systems generally fit into two categories: sensors in smart homes and smartphones. Smart home-based HAR relies on cameras placed around the household and pre-installed sensors to recognize specific activities. This method tends to be very accurate, but expensive and hard to install. On the other hand, smartphone-based HAR is mainly accelerometer-based -- using a user's movement patterns to identify activities. It is very easy to deploy since smartphones are ubiquitous, but inaccurate at detecting activities. 

On the classification front, smartphone-based systems are limited to identifying activities that are easily recognized from movement patterns, such as running, driving and sitting. Smart home designs typically are specialized and require additional hardware to identify any specific activity. Ideally, a system should classify a wide range of activities (cooking, brushing teeth, etc.) and even be able to identify progress or steps within an activity (e.g. mixing ingredients vs. cooking them). It should also be relatively easy to add new activities that the system can recognize.

Existing systems all suffer from some combination of inaccuracy, difficult deployment, and narrow range of recognized activities. 
The goal of this project is to see if it is possible to perform practical, high-accuracy HAR that can recognize a wide range of activities using only a smartphone. A key enabler for this system is improving smartphone sensor technology, which has resulted in smartphones having a wide variety of sensors. A particularly important addition in recent smartphones is hardware support for Wi-Fi Round Trip Time (also called Wi-Fi RTT or 802.11mc)~\cite{802.11mc}, which enables indoor localization with a precision of one to two meters. The addition of location information is particularly useful for HAR since many activities are performed in specific locations. For example, cooking is typically done only in the kitchen and brushing teeth is typically done in a bathroom. As a result, location information can help differentiate between activities that may seem similar when observed using other sensors, opening the potential for significantly improved accuracy. In fact, fine-grain localization at meter-level accuracy may help in recognizing different activities within an area, such as eating at the kitchen table, cooking at the stove or making coffee at a coffee machine. 

The Location-Enhanced HAR (\sys) system presented in this paper leverages acceleration, audio, and location data collected from a smartphone. The classification of these sensor readings into human activities relies on a neural network-based classifier. This neural network was trained using a labeled data set of sensor readings that I created for a set of twelve common activities (typing, vacuuming, washing dishes, running the blender, brushing teeth - electric, brushing teeth - regular, making coffee, taking medicine, using microwave, shaving, drying hair, and doing nothing). To better illustrate the value of adding new sensors, I also trained machine learning models using just the accelerometer readings and using a combination of accelerometer and microphone data. 


\sys achieved an $F_1$-score of 0.965 at recognizing twelve common activities described above. Using just acceleration, a $F_1$-score of 0.660 was achieved. Note that most existing smartphone systems use only acceleration for HAR. This proves that acceleration data is not very accurate in predicting activities similar to the ones I chose to test. Using only audio, a $F_1$-score of 0.865 was achieved, showing that \sys's use of location provides a more accurate approach to human activity recognition.


The rest of this paper is organized as follows. Section~\ref{sec:system} provides an overview of the \sys system, including the system requirements and the key techniques used. Sections~\ref{sec:collect},~\ref{sec:extraction}, and~\ref{sec:inference} focus on the design of the core software components of \sys. Section~\ref{sec:results} provides a detailed description of the results and an analysis of the data from this study. Discussion, Related Work and Conclusions are presented in Sections~\ref{sec:discussion},~\ref{sec:related} and~\ref{sec:conclusion}.


\begin{figure*}[t]
  \centering
  \includegraphics[width=.8 \textwidth]{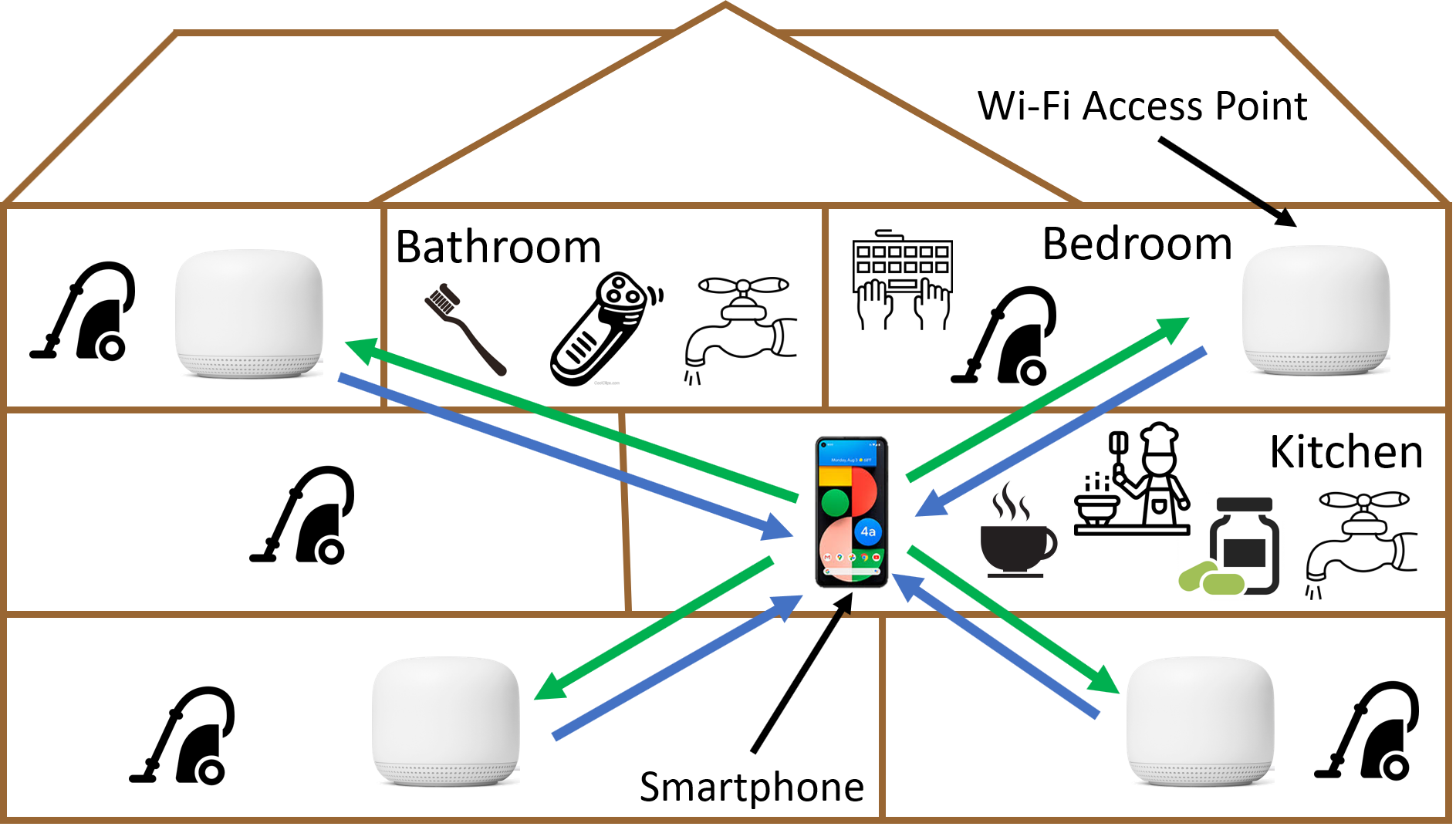}
  \caption{Diagram shows the location of the Google Nest devices throughout the home. The home used in the experiments had two floors and a basement. Two Nest access points were located on the upper floor and two were located in the basement. The figure also highlights the limited locations where particular activities were recorded. For example, making coffee and cooking were only observed in the kitchen, while brushing teeth was only observed in the bathroom. }
  \label{fig:home}
\end{figure*}

\section{System Overview}
\label{sec:system}

In this section, I describe the design of \sys. Section~\ref{sec:require} describes the key requirements that any system designed to perform HAR must address. Section~\ref{sec:hardware} provides an overview of the \sys hardware design and Section~\ref{sec:software} discusses the software design.

\subsection{System Requirements}
\label{sec:require}

The goal of this system is to develop a HAR system that is easy to deploy, accurately identifies activities and is easily extended to support new activities. This would enable a wide range of applications that need such activity context to provide useful information. For example, it could be used in applications that analyze the daily activity patterns of seniors to detect abnormal behavior or other health deterioration. 

In addition to meeting this high-level goal, the design of the \sys system needs to meet the following important requirements:

\begin{itemize}
    \item {\bf Easy Deployment.}  There should be no significant prior instrumentation in the household required. Many existing approaches to enabling HAR in homes require significant specialized infrastructure to be added around the house~\cite{2014.Krishnan}. 
    
    \item {\bf Portable.} Any sensors or components needed by the system must be either integrated into the mobile device or be similar in size/portability as the mobile device. Given the battery life constraints of mobile devices, the solution should use little if any power. 
    
    \item {\bf Unobtrusive.} The system should work without being intrusive to the user and not require the user to do anything for the system to operate.
    
    \item {\bf Accurate.} The system must have high accuracy. In order to support applications that depend on the recognition of multiple related activities, a HAR system must identify activities with an accuracy of at least 95\%.
    
    \item {\bf Inexpensive.} The system components must not add significant expense to the device.
    
    \item {\bf Extendable.} It should be easy to add new activities that the system can recognize. 
\end{itemize}

\sys employs a few main techniques to meet these requirements. First, it is a smartphone-only system; it relies only on the sensors that are available on a modern smartphone to collect data about a user's activity. This allows it to meet the Easy Deployments, Portable, Unobtrusive, and Inexpensive requirements. Second, it leverages a combination of different sensors, with fine-grain indoor localization as a key addition over any previous system, to recognize activities. This allows \sys to provide accurate predictions, meeting the Accurate requirement. Finally, \sys relies on a machine learning model to classify sensor observations to recognized activities. Retraining the machine learning model to recognize new activities can be easily automated, addressing the Extendable requirement. 

\subsection{\sys Hardware Overview}
\label{sec:hardware}

Given the smartphone-based design of \sys, the hardware consists solely of the smartphone itself and the communication infrastructure associated with it. Since my objective was to incorporate fine-grain location, I chose a smartphone and communication infrastructure that incorporates support for some of the newest localization techniques (802.11mc, aka. Wi-Fi RTT~\cite{802.11mc}). Google maintains a list of devices that support RTT-based localization on its Android Developer Guide~\cite{androiddev}. At the time of this project, a wide range of phones support this standard, including phones from Google, Samsung, LG, and Xiaomi. I choose a relatively inexpensive and easy to obtain phone, the Google Pixel 4a~\cite{pixel4a}. While phone support for the protocol is widespread, only Google-branded Wi-Fi access points support the protocol from an infrastructure perspective. I chose to use the Google Nest Wi-Fi Router and Google Nest Wi-Fi Point devices~\cite{nestwifi} for my infrastructure. 

Figure~\ref{fig:home} provides a pictorial view of the communication infrastructure within the home used for testing. The home has three floors -- a basement and two above ground floors. The Wi-Fi RTT protocol provides distance measurements to any access point that is within communication range (shown using the blue-green lines). Since Wi-Fi RTT only provides a distance estimate, measurements to four different access points is necessary to localize a device to a single location. This is explained in greater detail in Section~\ref{sec:tdoa}.
For this reason, I chose to deploy four access points in the home. In addition, I placed the access points at locations near the edge of the home, far from each other. This minimizes the localization error since the measurements to widely spaced access points produces a smaller intersection of potential locations.

\begin{figure*}[t]
  \centering
  \includegraphics[width=\textwidth]{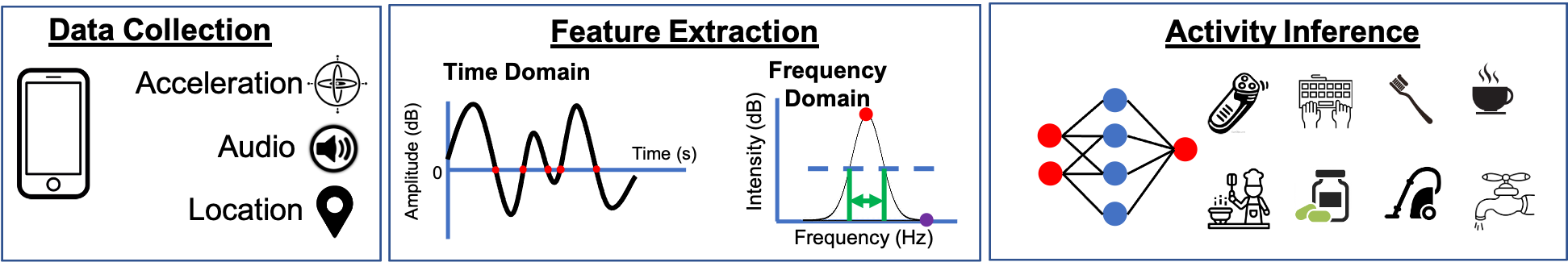}
  \caption{Diagram outlining the components of the \sys software.} 
  \label{fig:softwareDesign}
\end{figure*}

\subsection{\sys Software Overview}
\label{sec:software}
The overall structure of the \sys software is shown in Figure~\ref{fig:softwareDesign}. This software, which runs on the smartphone, consists of three main sections: data collection, feature extraction, and activity inference. The data collection software component of the system determines what data to collect and how often to collect it. I describe the software that performs this task in greater detail in Section~\ref{sec:collect}.
The feature extraction component of the code, described in Section~\ref{sec:extraction}, computes several defining features of the audio in the time and frequency domains. The goal of feature extraction is to process the sensor data into a format that is more applicable to machine learning. The activity inference component uses a trained neural network classifier to convert the raw sensor data and extracted features into recognized activities. The operation and training of this classifier is described in Section~\ref{sec:inference}.

\section{Data Collection}
\label{sec:collect}

One of the key design decisions in \sys is what data to collect. While modern smartphones have a wide range of sensors, some of them are not highly relevant to HAR and collecting excessive data can consume resources and reduce the battery life of the smartphone. Most deployed smartphone HAR systems collect only accelerometer readings. Unfortunately, to be effective, the recognized activities must have dramatically different motion properties. As a result, the range of activities that can be recognized using just acceleration data is relatively limited (just running, walking, sitting, etc.). Adding audio information can help broaden the set of activities since sounds associated with activities such cooking or brushing teeth can be distinctive. However, there are many activities that may have similar environmental sounds, such as those associated with small motors (e.g. electric toothbrush, microwave oven, blender). In addition, audio features may be error prone since the environment may be noisy or the smartphone may not be optimally located to record sound (e.g. inside a pocket). To further distinguish between activities, \sys also collects location information. As shown in Figure~\ref{fig:home}, many activities are location specific; for example, tooth brushing typically only occurs in the bathroom. \sys is designed on the premise that the combination of acceleration, audio and location information suffice for performing highly accurate HAR. This section describes what data \sys collects and how it implements this data collection. 

\subsection{Indoor Localization}

The most common approach to obtaining location information on mobile devices is to rely on GPS (Global Positioning System~\cite{GPS}). While GPS is relatively accurate, GPS signals do not propagate through walls and, as a result, GPS cannot provide location information indoors. Unfortunately, there are no widely deployed standards for getting location information indoors. Existing systems use one of three basic techniques: angle of arrival, signal strength, and time difference of arrival.

\subsubsection{Angle of Arrival Localization} 
~

Angle of arrival localization systems~\cite{aoa1,aoa2} work by measuring the angle that a transmitter's signal arrives at a user's device. If there are multiple access points in a home, the user's device could observe the angle to each of the different access points. If the location of the access points is known, the device can then compute its own location from this angle information. Unfortunately, measuring the angle of arrival requires that the mobile device have some form of directional antenna. These are typically large and impractical for something like a smartphone. This leaves me two other alternatives that I consider below. 

\subsubsection{Signal Strength Localization} 
\label{sec:rssi}
~

Signal strength based localization systems (e.g.~\cite{RADAR}) leverage the principle that Wi-Fi signals typically get weaker further away from a transmitter. In an ideal setting, the signal strength would vary inversely with the square of the distance; however, in practice, there are many other factors beyond just distance that impact signal strength at a location. For example, signals typically attenuate significantly as they pass through obstacles such as walls. As a result, walls between a transmitter and a user may make the signal much weaker and seem like the transmitter is further away. Wi-Fi signals also suffer from multi-path interference. When a transmitter sends a message, it is sent out in all directions. The signal that travels directly between the transmitter and the receiver may interfere with the signal that reflects off a surface and then propagates to the receiver. This interference can be constructive or destructive (increasing the signal strength or reducing the signal strength, respectively) depending on the difference in distance between the direct and indirect path. The end result of these factors is that signal strength only provides a rough approximation for distance. 

Practical systems that use signal strength for localization~\cite{RADAR} rely on someone generating a map of signal strength to location for a building for every single access point in the building. Surveying a building to generate this map usually takes significant effort and impedes wide use of this approach. Devices that want to determine their location measure the signal strength from all nearby access points and then lookup these signal strengths on the signal strength map. Even with detailed maps of signal strength, the accuracy of these systems are relatively low, allowing users to localize themselves to only a 3-4 meter region.

\begin{figure}[t]
  \centering
  \begin{tabular}{c}
      \includegraphics[width=.75\linewidth]{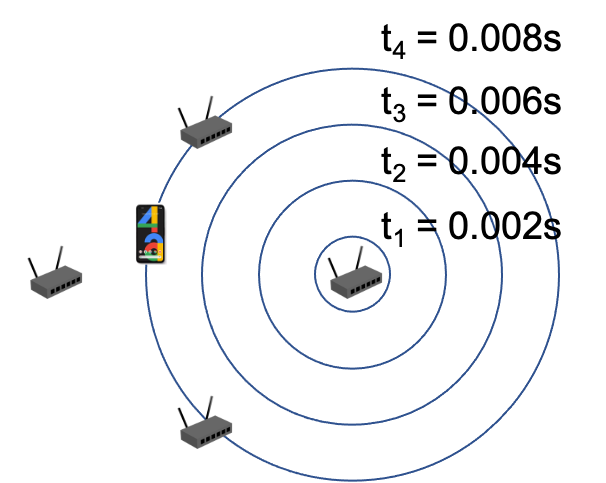} \\  
       (a) \\
        \\
         \includegraphics[width=.75\linewidth]{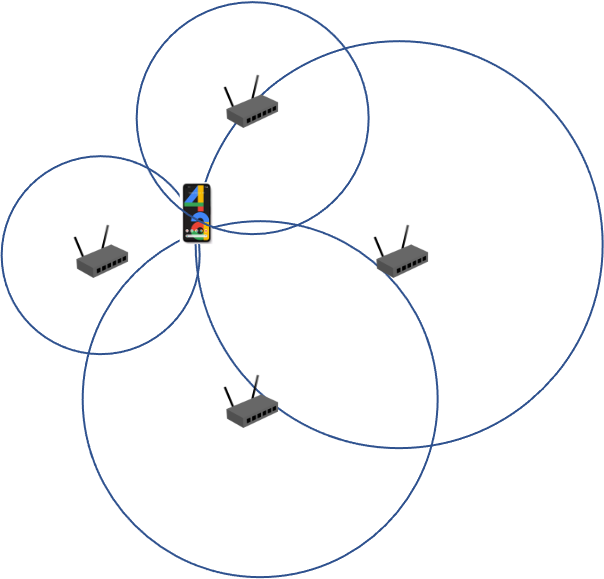}  \\
       (b) \\
       \\
  \end{tabular}

  \caption{Illustration of TDoA localization. (a) shows the time for a signal to propagate and (b) shows multilateration used to localize the device from multiple distance measurements.}
  \label{fig:tdoa}
\end{figure}

\begin{figure}[t]
  \centering
  \includegraphics[width=0.75\linewidth]{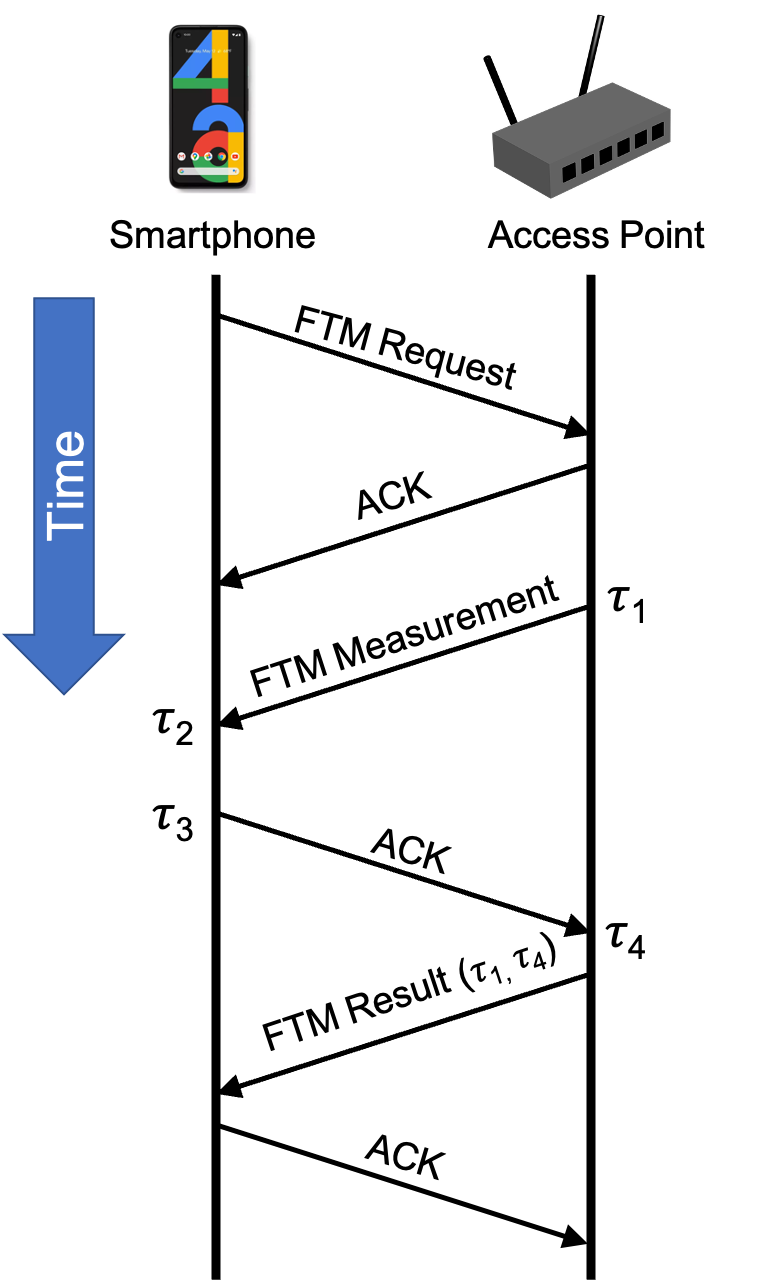}
  \caption{Wi-Fi RTT measurement protocol message exchange.}
  \label{fig:wifirtt}
\end{figure}

\subsubsection{Time Difference of Arrival Localization}
\label{sec:tdoa}
~

Figure~\ref{fig:tdoa} provides a 2-dimensional illustration of the concepts behind Time Difference of Arrival (TDoA) localization.
TDoA leverage the observation that signals take time to travel between a transmitter and receiver. This is shown in Figure~\ref{fig:tdoa}(a) where the signal propagates outward from the access point on the right, reaching the smartphone at $t_4 = 0.008s$. If a device can measure the time it takes for the signal to travel, it can use the propagation speed of the signal and the measured time to compute a distance to the transmitter. This narrows the possible locations of the receiver to a spherical shell at the measured distance from the transmitter. In 2-dimensions, this would represent a circle instead of a sphere. The circles around the access points in Figure~\ref{fig:tdoa}(b) show the potential locations of the smartphone based on the distance measurements to the respective access points. If this measurement is performed with two different transmitters, the receiver's location can be narrowed further to the intersection of the two associated spherical shells -- this typical produces a circle of possible locations. Intersecting this circle with the spherical shell from another measurement results in a pair of points where the receiver may be located. Collecting measurements from four transmitters narrows the potential location of the receiver to a single spot. In the 2-dimensional example shown, only three access points are needed. This technique has been used since World War II in navigation systems~\cite{williams2003loran} and is used in systems such as the Global Positioning System (GPS)~\cite{GPS}.  Note that GPS systems rely on TDoA measurements from satellites. 

A key challenge in TDoA systems is actually measuring the propagation time. Imagine a simple strawman design in which an access point reads its clock and transmits the local time in a packet. The receiver could then simply read the clock when the message is received and subtract the local time from the time in the packet. This would seem to provide a simple way to determine the time it took for the message to travel from the access point to the receiver. However, there are three major flaws with this approach. 

The first issue is with clock synchronization. The clock at the access point and the clock at the receiver may not be synchronized, and they may read different times at the same moment. In addition, one clock may run faster than the other and they drift further and further ahead over time. Any measurement of the propagation time would have this additional offset (either positive or negative) from this lack of synchronization. This can be addressed by measuring the propagation time in both directions (also called a round trip time) and dividing by two. This would incorporate both a positive and negative offset, which would cancel each other out in the measurement. 

The second issue is that the clocks need to be very precise. Wi-Fi signals propagate at the speed of light ($3.0 \times 10^8 m/s$). This means that the signal travels 1 meter in 3.34 nanoseconds. 
If a clock only has a precision of $1\mu s$, it would only be able to measure distances $\pm 300m$. Extra hardware support is needed to provide clock readings with nanosecond precision. 

The third issue is processing time. Computers process data and messages at a finite speed. The processing time will get added into the propagation time measurements. Hardware support to timestamp messages as the signal arrives is needed to obtain meter level accuracy in measurements. 

Wi-Fi added support for TDoA localization to the standard in 2016~\cite{802.11mc}. This standard, called 802.11mc or Wi-Fi Round Trip Time (Wi-Fi RTT), enables smartphones to determine the distance from access points with a precision of 1-2 meters based on time it takes signal to travel to device and back. 

The basic message exchange used in Wi-Fi RTT is shown in Figure~\ref{fig:wifirtt}. The smartphone normally scans for nearby access points to associate with for network connectivity. As part of this scan, the smartphone learns about the capabilities of the nearby access points including whether they support the Wi-Fi RTT protocol. When a smartphone wishes to measure the distance to an access point that supports Wi-Fi RTT, it initiates a ranging request by transmitting a Fine Timing Measurement (FTM) request to the access point. The access point immediately acknowledges receipt of this request by sending an ACK response and scheduling a measurement exchange. At some later time ($\tau_1$), the access point transmits an FTM Measurement packet. This arrives at the smartphone whose Wi-Fi interface records time $\tau_2$. The measurement packet is processed by the smartphone and an acknowledgement (ACK) is transmitted at time $\tau_3$. The access point records the time ($\tau_4$) when this ACK message is received. It then transmits the times $\tau_1$ and $\tau_4$ to the smartphone.

Note that the time $\tau_1$ and $\tau_4$ are recorded using the access point's clock and times $\tau_2$ and $\tau_3$ are recorded using the smartphones clock. These clocks are not synchronized in any way. Despite this, the round trip propagation time for the messages can still be calculated as:

\begin{equation} \label{eq:1}
Round~~Trip~~Time = (\tau_4 - \tau_1) - (\tau_3 - \tau_2)
\end{equation}

\noindent Note that the first term ($\tau_4 - \tau_1$) only uses clock values from the access point and the second term ($\tau_3 - \tau_2$) only uses times from the smartphone. As a result, the synchronization issues do not impact this computation in any way. With this measurement complete, the distance to the access point can be computed by the smartphone as: 

\begin{equation} \label{eq:2}
Distance = c \times \frac{Round~~Trip~~Time}{2}  
\end{equation}

\noindent
Note that $c =$ the speed of light = $3.00 \times 10^8 m/s$.\\
The $Round~~Trip~~Time$ is divided by two to get the one-way propagation delay for the signal.

\begin{figure*}[t]
  \centering
  \includegraphics[width=.75\textwidth]{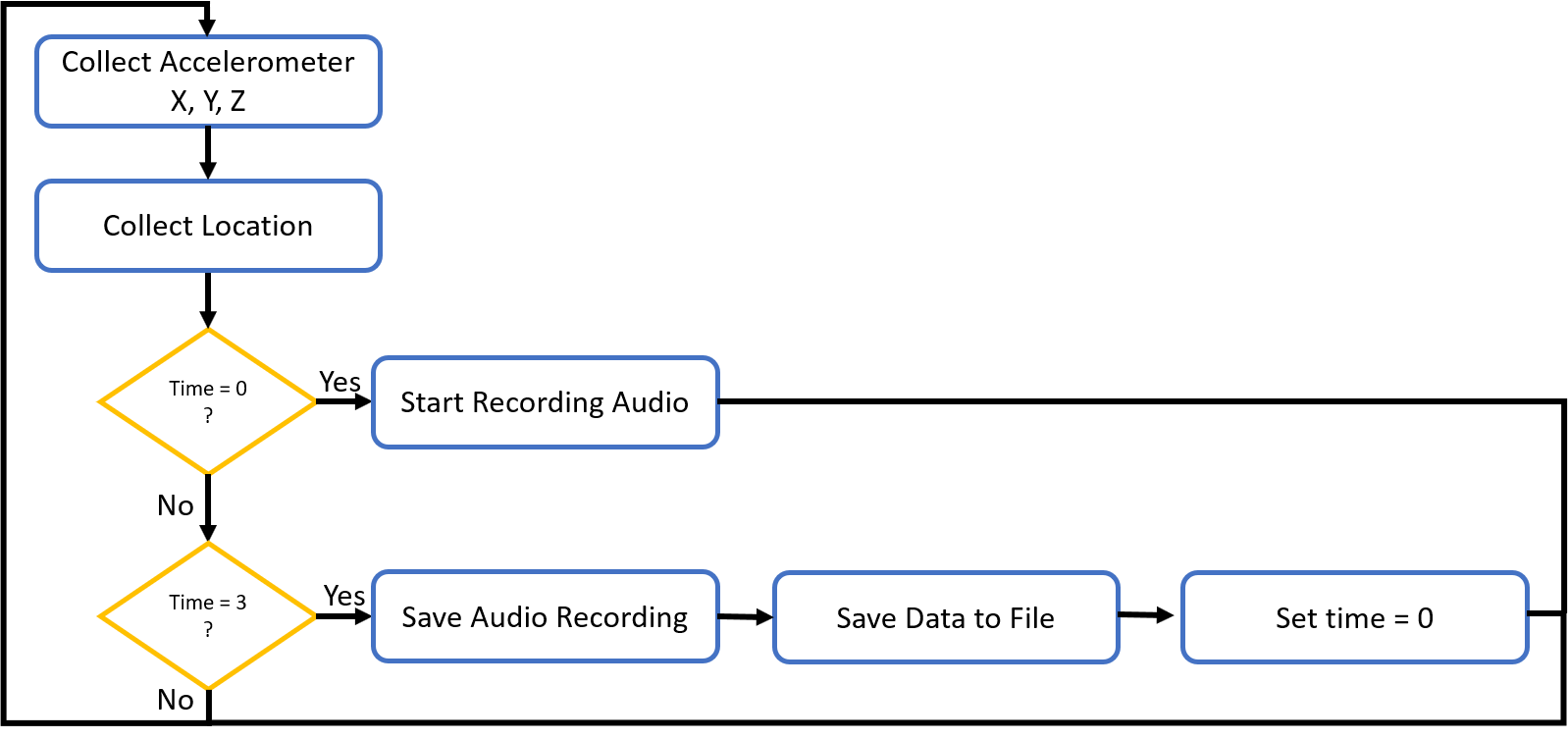}
  \caption{Flowchart representing the high level software structure for data collection android application}
  \label{fig:flowchart}
\end{figure*}

\subsubsection{Discussion}
~

The first observation to notice is that the design of \sys does not exactly require the exact location of the user in physical space. What \sys requires is a set of sensor readings related to location that are closely correlated to the possible set of activities performed. The reason for this is that a machine learning based classifier is used to convert the sensor readings into recognized activities. This simplifies the collection of location information since \sys does not require three dimensional coordinates for an end user. \sys could just as easily use the array of distances to different access points as a feature in its machine learning classifier. 

The second observation is that Wi-Fi RTT is a much better choice for \sys than a signal strength based approach due to its much higher accuracy. Measurement studies of the accuracy of Wi-Fi RTT~\cite{wifirtt_accuracy} have shown that the variation in distance estimates for the same location is approximately 1 meter. This level of accuracy would allow \sys to differentiate between activities within a single room. For example, it could allow \sys to determine the difference between making coffee (near the coffee machine) from using the microwave (near the microwave oven) as long as the coffee machine and microwave oven are not in the exact same location. 

\begin{figure}[t]
  \centering
  \begin{tabular}{c}
      \includegraphics[width=\linewidth]{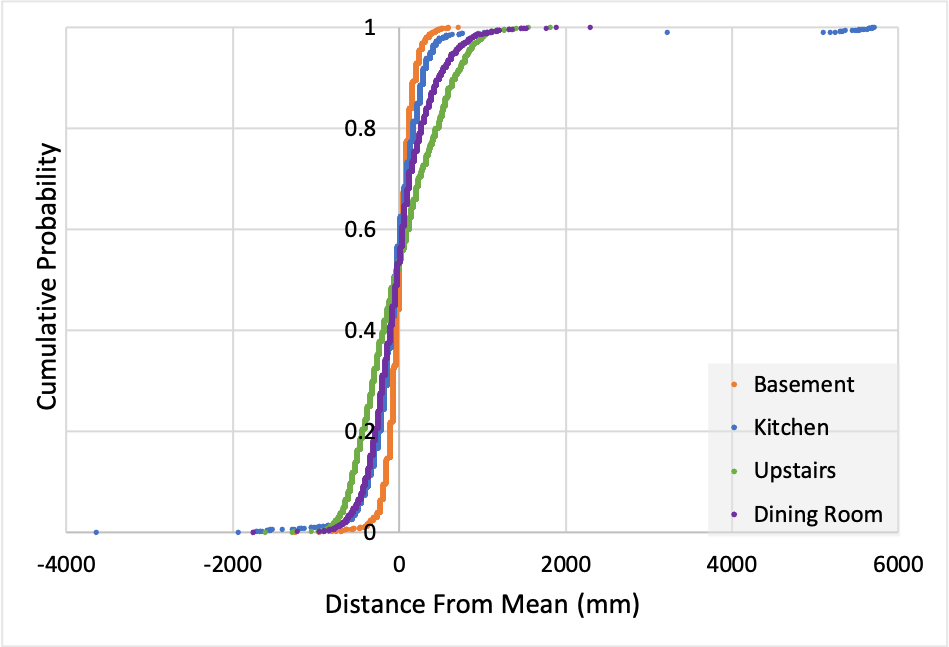} \\  
       (a) \\
        \\
         \includegraphics[width=\linewidth]{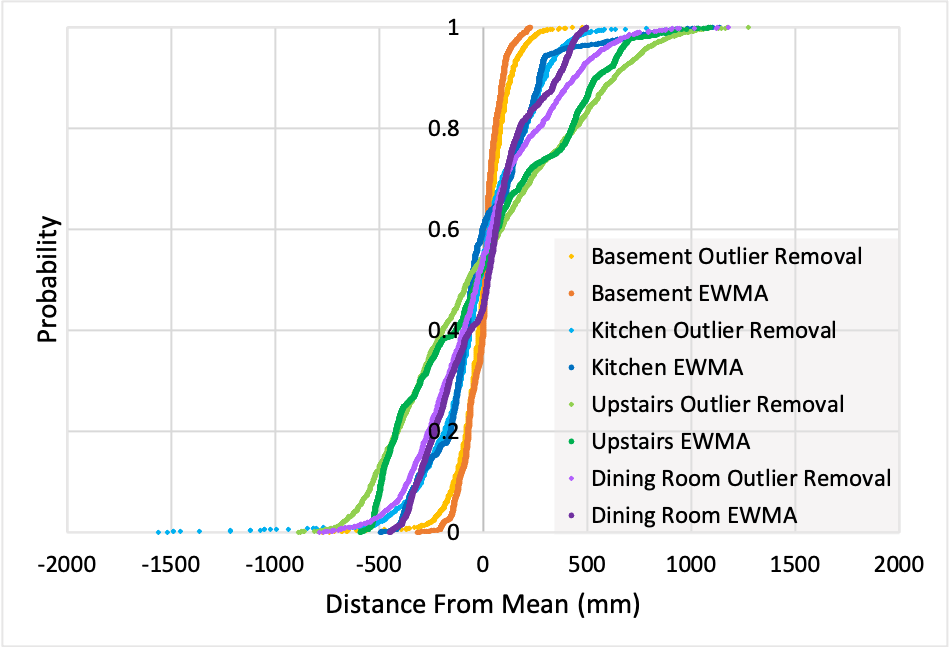}  \\
       (b) \\
       \\
  \end{tabular}

  \caption{Measurement of Wi-Fi RTT accuracy. (a) shows the raw measurements for distance to the access points. The distances are plotted as a CDF of difference from the mean distance and (b) shows distance measurements passed through different filters. EWMA lines represents an exponentially weighted moving average of the readings and outlier removal lines presents the results of filtering based on the range of recent measurements.}
  \label{fig:wifirttmeas}
\end{figure}

To better understand the potential of Wi-Fi RTT, I performed a measurement study of its accuracy in the home shown in Figure~\ref{fig:home}. While leaving the phone at a fixed location, I recorded the distances measured to all four access points. 2286 distance measurements were collected to each access point. For each access point, I compute the mean distance measured across all its measurements. In Figure~\ref{fig:wifirttmeas}(a), I plot a cumulative distribution function (CDF) of each measurement minus the mean distance to that access point. Ideally, the measurements to a single access point would all be identical since the phone was not moving and the CDF would be a sharp step function at 0. However, in practice, the measurements show some variation. The distribution of values around 0 appear symmetric and roughly normally distributed. The standard deviation of each of the access points differs slightly from each other with the kitchen access point having the worst standard deviation of 630 mm. At first glance, this would seem to match well with the reported accuracy of 1 to 2 meters. However, note that in Figure~\ref{fig:wifirttmeas} there are a number of outlier measurements that are between 4000 and 6000 mm from the mean. This variation in measurements would make it difficult to use the distance measurements effectively for activity recognition. I tested a few different algorithms to remove these outliers. First, I tested a simple outlier removal that looked at the range of the last 5 readings. If the range was less than 1000mm, the algorithm simply reported the current value. If the range exceeded 1000mm, the algorithm removed the largest and smallest reading of the 5 and reported the average of the remaining 3 readings. I also tested an exponentially weighted moving average (EWMA), where a new average value would be calculated when a new reading was made as:
\begin{equation}
average_{new} = \alpha \times reading_{new} + (1 - \alpha) \times average_{old}   
\end{equation}
I used an EWMA with $\alpha = 0.1$. This EWMA acts as a low-pass filter eliminating spurious spikes/dips in the readings. The results of using these two algorithms is shown in Figure~\ref{fig:wifirttmeas}(b). As can be seen from the graph, the spread of readings is significantly reduced and outliers are eliminated. The the outlier removal and EWMA on the kitchen access point measurements results in a standard deviation of 368 and 243mm, respectively. In general, the EWMA produced narrower distributions and I use this for all my experiments.

\subsection{Implementation}

The Flowchart in Figure~\ref{fig:flowchart} provides a high-level view of the code used to collect the needed sensor readings from an android smartphone. The Java app continuously collects x, y, and z acceleration data as well as the Wi-Fi RTT distances from the 4 routers. Every 3 seconds, it saves an audio file and starts recording a new clip. Additionally, it saves the acceleration and location data to a file.

In order to collect acceleration data, the code creates a SensorManager object, which monitors the activity of the accelerometer. When an update to the acceleration is received, it records the data to an array. The absolute value of the acceleration data is averaged across a 3 second period.

The next step in the code is to collect the location data. To do that it needs to first find all of the access points in range that support Wi-Fi RTT. It does this by using the WifiManager class to interact with Wi-Fi access points. The code records the BSSID for each of the access points. BSSID stands for basic service set identifier and is a unique address for each access point.
The code then creates a WifiRttMananger object to measure the distance in millimeters to each of the 4 access points. It does this by defining and executing a ranging request to receive the RTT information. 

The average acceleration reading and the Wi-Fi RTT data is saved to a new line in a csv file once per 3 seconds.

The last piece of information the code needs to collect is audio. It creates a MediaRecorder object and saves audio clips as a wav file every 3 seconds, synchronized with the collection of the other sensor readings.

\section{Feature Extraction}
\label{sec:extraction}

While the accelerometer and Wi-Fi RTT readings are simple scalar values that are easily used in a machine learning classifier, the audio signal is a more complex data type. 
In order to classify the audio signals, the system must extract simple features that are unique to the particular sound. \sys uses seven features of the audio, extracted on both the time and frequency domain.

\subsection{Time Domain Features}

The time domain features used by \sys focus on the amplitude and shape of the audio signal. 

The first feature, root-mean-square-energy, focuses on the raw amplitude of the audio signal. The energy of a signal is the sum of the magnitudes of the signal squared. The root mean-square-energy is then the square root of the mean energy. It provides a useful signature of the loudness of the recorded sound. 

\begin{figure}[t]
   \centering
   \includegraphics[width=\linewidth]{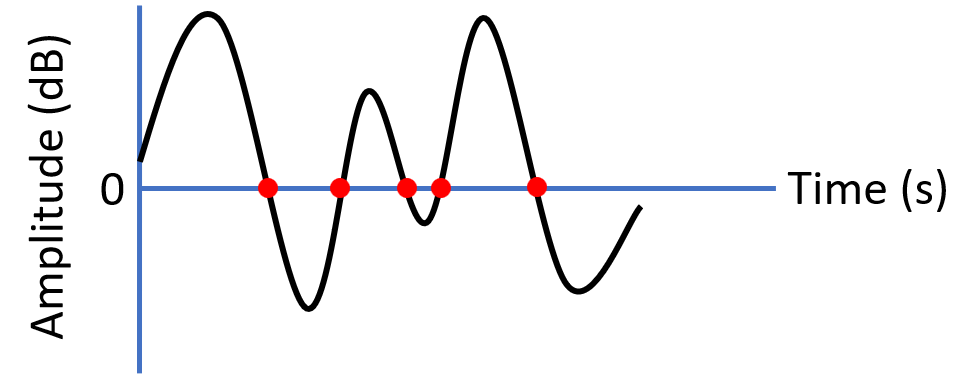}
   \caption{This graph shows what the zero crossing rate represents. The feature is computed on a time domain meaning that amplitude is on the y-axis and time is on the x-axis. Zero crossing rate is the number of times the signal amplitude crosses zero.}
   \label{fig:zerocrossing}
 \end{figure}

The second feature, zero crossing rate, summarizes the shape of the signal. The computation of the zero crossing rate is illustrated in Figure~\ref{fig:zerocrossing}. The zero crossing rate is the number of times an audio signal crosses zero, shown the red dots shown in the figure. Past work~\cite{zerocrossing} has shown that this value is useful for distinguishing between a range of naturally occurring sounds.

  \begin{figure}[t]
   \centering
   \includegraphics[width=\linewidth]{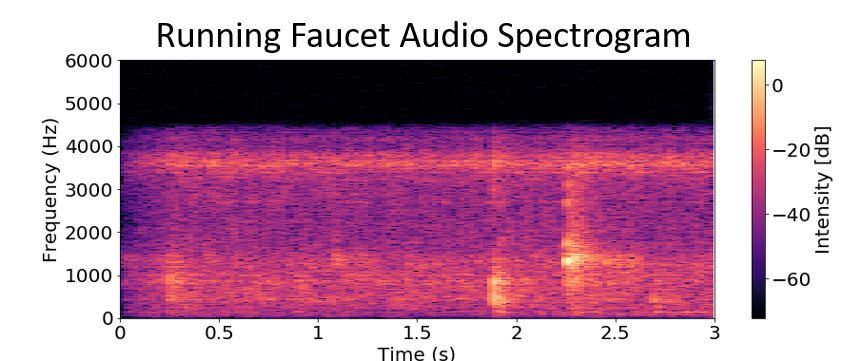}
   \caption{This graph describes the energy level at each frequency over time in an audio recording of a running faucet.}
   \label{fig:faucetSpec}
 \end{figure}
 
   \begin{figure}[t]
   \centering
   \includegraphics[width=\linewidth]{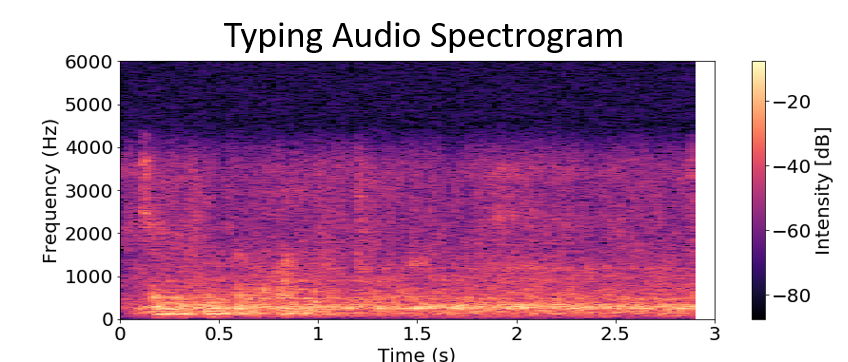}
   \caption{This graph describes the energy level at each frequency over time in an audio recording of typing.}
   \label{fig:typingSpec}
 \end{figure}

 \begin{figure}[t]
   \centering
   \includegraphics[width=\linewidth]{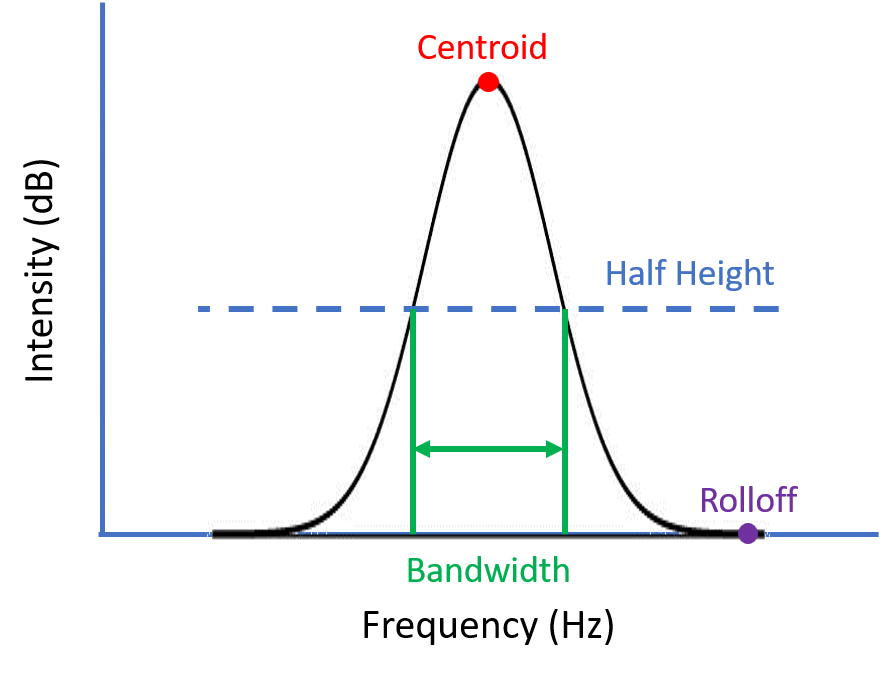}
   \caption{This graph shows the features that can be extracted on a frequency domain, meaning that intensity is on the y-axis and frequency is on the x-axis.}
   \label{fig:freqDomain}
 \end{figure}

\subsection{Frequency Domain Features}

The frequency domain features used by \sys focus on how the energy of the audio signal is distributed across different sound frequencies. 
The first step in obtaining any of these features is to perform a Fourier transform~\cite{fourier} on the audio signal. This converts the time domain samples to an equivalent set of frequencies and amplitudes. From this frequency domain representation, \sys computes five different features.

For \sys's first frequency domain feature, it uses a variant of the Fourier transform called short-time Fourier transform or STFT~\cite{STFT}. This variant is particularly useful for audio signals since it provides frequency content of local sections of the audio as it changes over time. 
Spectrograms, such as the ones in Figure ~\ref{fig:faucetSpec} and Figure ~\ref{fig:typingSpec} are a visual representation of STFT output. The frequency is on the y-axis, time on the x-axis, and the brighter the color, the greater the amplitude (i.e. intensity) of the audio signal at that frequency. Notice the thin horizontal bar of brighter color in both figures. This is because there was a greater intensity at that frequency in the audio clip. For example, in Figure~\ref{fig:faucetSpec}, the horizontal band at 3700Hz is a unique characteristic of this faucet's noise and the vertical lines near 1.9 and 2.3 seconds represent short bursts of additional wide-frequency noise, possibly created by a background event or splashing of water. The entire set of coefficients are produced from the STFT of the 3 second audio clip and the mean is used as a feature.


The remaining frequency domain features attempt to summarize the shape of the audio signal in frequency domain. Figure~\ref{fig:freqDomain} shows three of these features: centroid, rolloff and bandwidth. Spectral centroid describes the frequency that the energy is centered on. The spectral rolloff is the frequency that high frequencies go down to 0. The spectral bandwidth is the range of the frequencies with significant intensity. The final feature is the  Mel-Frequency Cepstral Coefficients~\cite{MFCC} of the audio signal. The MFCC measure is similar to the STFT coefficients -- however, the MFCC values are taken from the frequency bands on the mel scale, which are more representative of human hearing of sounds. As a result, it is often especially useful in speech recognition or other human-generated sounds. 


\begin{figure}[t]
   \centering
   \includegraphics[width=.75\linewidth]{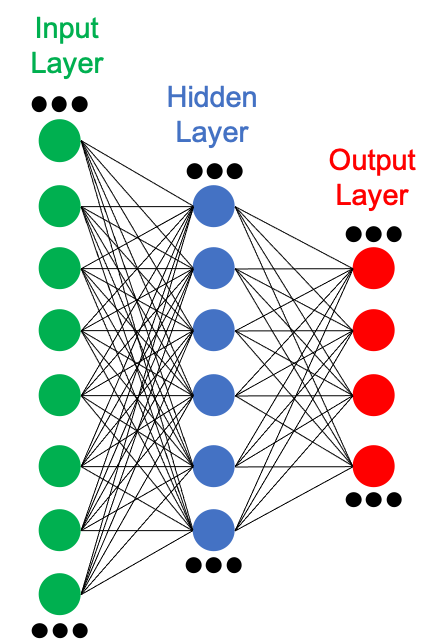}
   \caption{This diagram represents how the neural network is structured.}
   \label{fig:neuralNet}
 \end{figure}

\section{Activity Inference}
\label{sec:inference}



In order to identify the activity being performed from the raw sensor readings and extracted features, \sys uses a neural network. Neural networks are a type of supervised machine learning most commonly used for such classification tasks. Like any other supervised machine learning system, a neural network needs data labeled with the ground truth, such as a set of sensor readings labeled with the associated activities. Section~\ref{sec:data} describes the data set collected for this purpose. Once this data is collected, I also need to determine the structure of the neural network to use (Section~\ref{sec:structure} and train this neural network using the data collected (Section~\ref{sec:training}).



\subsection{Data Collection}
\label{sec:data}

As described in Section~\ref{sec:collect}, \sys uses x,y, and z acceleration, four distance values (one from each access point), and around 1500 three second audio clips collected on a smartphone to identify activities. To train \sys's neural network, I collected these sensor readings.

Data was collected for twelve common activities (typing, vacuuming, washing dishes, running the blender, brushing teeth - electric, brushing teeth - regular, making coffee, taking medicine, using microwave, shaving, drying hair, and doing nothing). The activities were specifically chosen because they are difficult to detect using existing smartphone HAR methods. Data was manually collected by running the smartphone application and performing the activities while the data was saved. 

The data was then manually labelled with the performed activity. Labelling was done by placing the audio into a folder with the name of the activity. The code reads the name of the folder and appends it to a text file with the acceleration and location data, along with the seven audio features described in Section~\ref{sec:extraction}.

Before training the model, the data is split into train and test data in order to evaluate the model accuracy on data it has not been trained on. 80\% of the acceleration, audio, and location data is used to train the data while the other 20\% is set aside for testing the \sys system. 

In addition, the data is scaled using StandardScalar() resulting in each feature having a distribution with a standard deviation of one. The reason for scaling is that features that have a larger range tend to hold a greater significance in training the model compared to features with smaller ranges. Scaling each feature to have the same standard deviation normalizes the data and removes this bias.

\subsection{Neural Network Structure}
\label{sec:structure}

Figure ~\ref{fig:neuralNet} shows the structure of the trained neural network. The colored dots represent artificial neurons that are connected to each other in a network. The neurons are separated into layers where the first layer (highlighted in green) is the input layer, the middle layers (highlighted in blue) are hidden layers, and the last layer (highlighted in red) is the output layer.

Each neuron computes a weighted sum of the input and adds on a bias. Each added layer adds a level of complexity in the decision making process of the model. This is because the neurons in a layer weigh the results of the previous layer to make a decision. After several tests with different number of layers, three layers seemed sufficient in producing satisfactory results.

Each of the artificial neurons in the network include an activation function. An activation function determines what should be fired to the next neuron in the network. It takes input data and produces an output. There are a variety of functions that can be used when designing a neural network and each layer can be assigned a different activation function for its neurons ~\cite{kerasapi}. This project uses four Rectified Linear Unit (ReLU)~\cite{relu} and two Softmax~\cite{generalML} functions. Based on testing with different activation functions, this combination worked the best. ReLU and Softmax are very commonly used activation functions in machine learning. In the ReLU activation function, the function f(z) is equal to zero if z is less than zero. If z is greater than or equal to 0, f(z) is equal to z. It is defined by the following equation:
\begin{equation} \label{eq:3}
{\displaystyle \phi (\mathbf {v} )=\max(0,a+\mathbf {v} '\mathbf {b} )},
\end{equation}
The Softmax function is used in the last layer to normalize the output of the neural network to a probability distribution. It is defined by the following equation:
\begin{equation} \label{eq:4}
\sigma(\vec{z})_{i}=\frac{e^{z_{i}}}{\sum_{j=1}^{K} e^{z_{j}}}
\end{equation} 

In the code, the input layer is defined with an input shape, which defines the shape of the starting tensor. The input shape depends on the shape of the test data (depending on the amount of data and the number of features). The input layer has 2430 neurons that the input data is fed into. The output layer is where the model outputs what activity is being performed. Thus, there are twelve neurons in the output layer. 

In addition to training a model using all of the features, I trained two more models. One of them was trained solely on the x, y, and z acceleration data, and one was solely trained on the audio features. The goal was to see what impact Wi-Fi RTT technology had on the \sys system. The next section describes the results of the model training.

\subsection{Training}
\label{sec:training}

\begin{figure}[t]
   \centering
   \includegraphics[width=\linewidth]{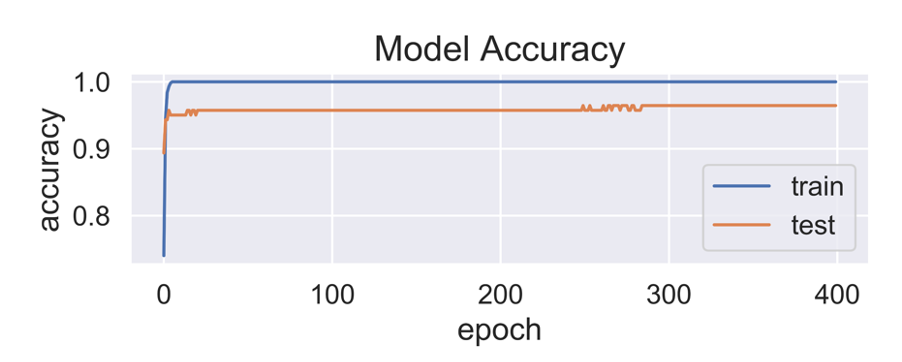}
   (a)
   \centering
   \includegraphics[width=\linewidth]{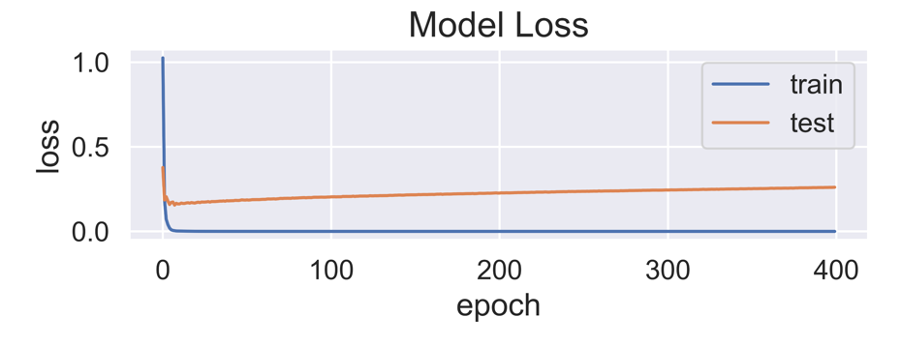}
   (b)
   \caption{(a) shows the accuracy and (b) shows the loss of the model as it is being trained.}
   \label{fig:loss}
 \end{figure}

Code was written to create a neural network of the appropriate structure and train it. The Blue (train) line in Figure~\ref{fig:loss} shows the accuracy and loss when training the neural network at each iteration of optimization (epoch).  Accuracy is the percentage of the predictions it got correct out of the total and loss describes how far away a prediction was from the actual value. At the end of training, the training accuracy and loss have leveled out suggesting that additional training time will not help. The accuracy on the training data set reaches 100\% at the end of training.

\section{Results}
\label{sec:results}

In my experimental evaluation of \sys, I consider three key questions:
\begin{enumerate}
    \item How accurate is \sys at identifying activities and is it signficantly more accurate than existing approaches? (Section~\ref{sec:accuracy})
    
    \item How does the addition of more recognized activities impact the accuracy of \sys? (Section~\ref{sec:scaling})
    
    \item How much and how accurate does location information need to be to provide the benefits of the \sys system? (Section~\ref{sec:locVacc})
\end{enumerate}

\subsection{\sys Accuracy}
\label{sec:accuracy}

To understand how well \sys performs, I tested \sys on the 20\% of data set aside for testing and validation (see Section~\ref{sec:data}). Figure~\ref{fig:loss} displays the accuracy and loss of the neural network at each iteration of training (epoch) on this test data (the red test line).
\sys achieved a validation accuracy of 96.5\% (accuracy when attempting to predict the activities in the testing data).

  \begin{figure}[t]
   \centering
   \includegraphics[width=\linewidth]{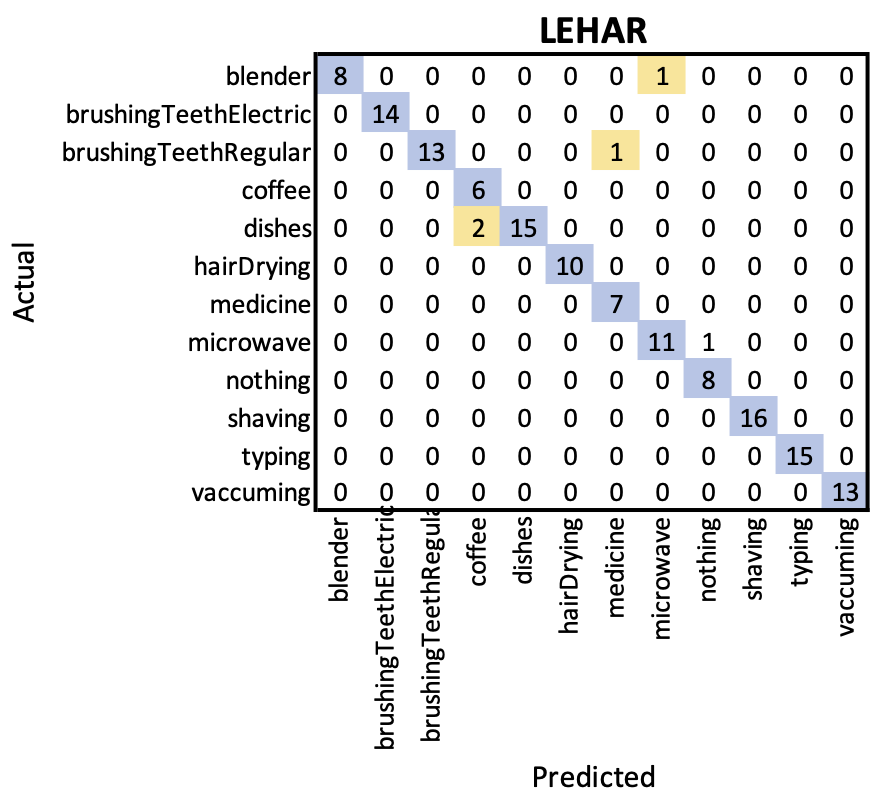}
   \caption{This figure describes the counts for each prediction of an element in the test data by the model and the actual label for the \sys system.}
   \label{fig:LEHARconfusionMatrix}
 \end{figure}
 
   \begin{figure}[t]
   \centering
   \includegraphics[width=\linewidth]{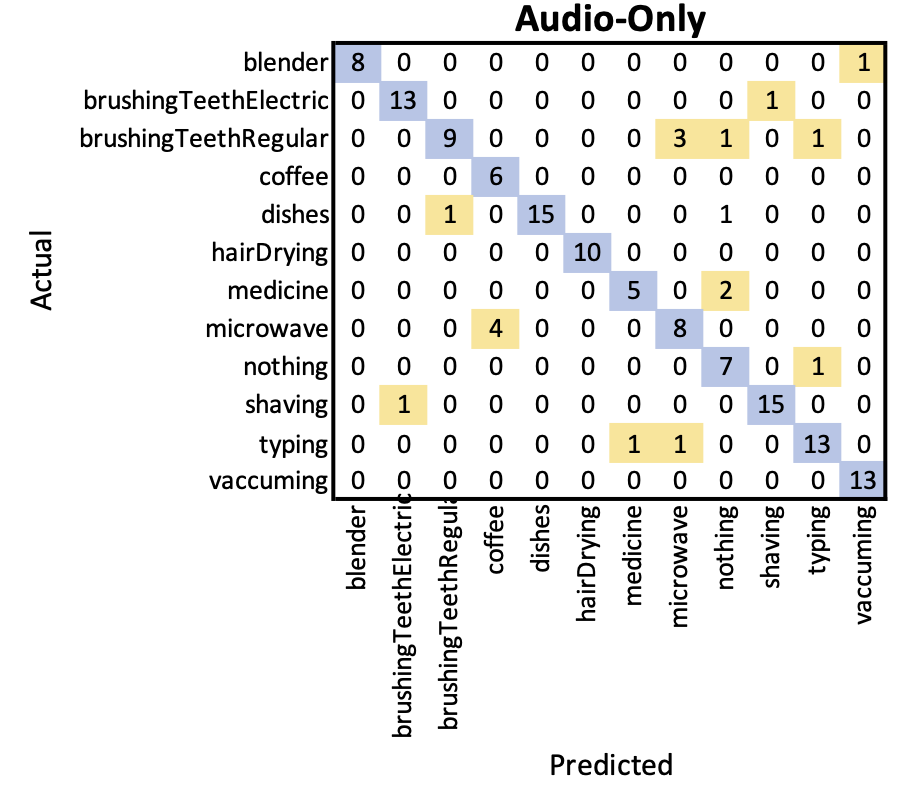}
   \caption{This figure describes the counts for each prediction of an element in the test data by the model and the actual label only using audio data.}
   \label{fig:audioconfusionMatrix}
 \end{figure}
 
 \begin{figure}[t]
   \centering
   \includegraphics[width=\linewidth]{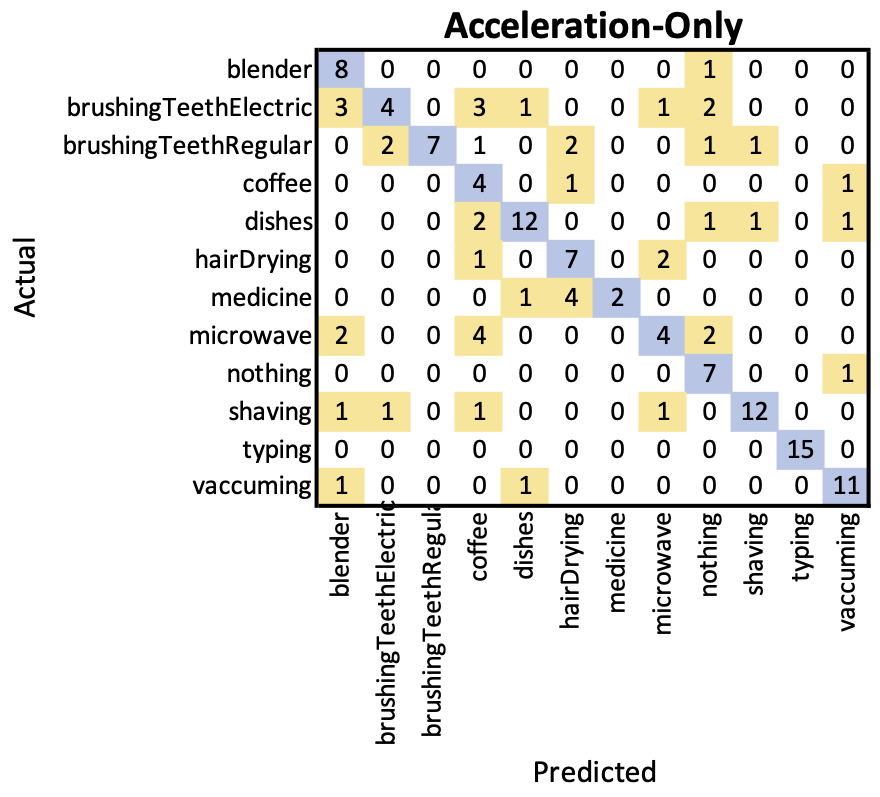}
   \caption{This figure describes the counts for each prediction of an element in the test data by the model and the actual label only using acceleration data.}
   \label{fig:accconfusionMatrix}
 \end{figure}
 
To better understand the nature of the errors made by \sys in recognizing activities, I created a confusion matrix for its activity predictions (Figure~\ref{fig:LEHARconfusionMatrix}). The rows represent what the true activity was and the columns represent what the model predicted as the activity. The boxes contain the count of such predictions/reality pairs. The boxes along the diagonal (highlighted in blue) represent situation where the actual activity and predicted activity match (where the model correctly guessed the performed activity). The yellow boxes indicate situations where the model incorrectly identified an activity. As indicated by the high counts in the blue boxes compared to the yellow, the model performed relatively well. The overall $F_1$-score for the LEHAR model was 0.965. There is no clear, statistically significant pattern to the errors to suggest a common source of error. 

Figure ~\ref{fig:audioconfusionMatrix} and ~\ref{fig:accconfusionMatrix} are confusion matrices for the audio-only and acceleration-only models. Note that the audio-only model made several incorrect predictions by predicting an audio clip of using the microwave as making coffee. This is most likely because making coffee and using the microwave are similar sounding activities. The \sys system did not make the same mistake as often because it includes location data and using the microwave and making coffee do not occur in the same location withinin the kitchen. The acceleration-only model made incorrect predictions much more often than the other two models. Activities such as typing are identified reliably since it is the only activity in the list that is performed while sitting\footnote{While sitting, the phone is in consistently in a different orientation since the phone was in my pant pocket}. 

The model trained solely on acceleration data produced a an $F_1$-score of 0.660 . The low accuracy indicates that most current HAR systems that rely on solely acceleration information would not be capable of identifying complex activities such as the ones chosen in this project. This explains why current smartphone HAR systems only focus on movement based activities such as running and walking.

The model trained solely on the audio features achieved an $F_1$-score of 0.865. The use of audio data is an important factor in accurately predicting the activities. In addition, the confusion between making coffee and using the microwave suggests that other similar sounding activities will cause issues as more and more activities are added to the system. \sys's combined model performed significantly better, proving that the use of WI-Fi RTT and the use of multiple smartphone sensors enable an extremely high accuracy HAR system.

 \begin{figure}[t]
   \centering
   \includegraphics[width=\linewidth]{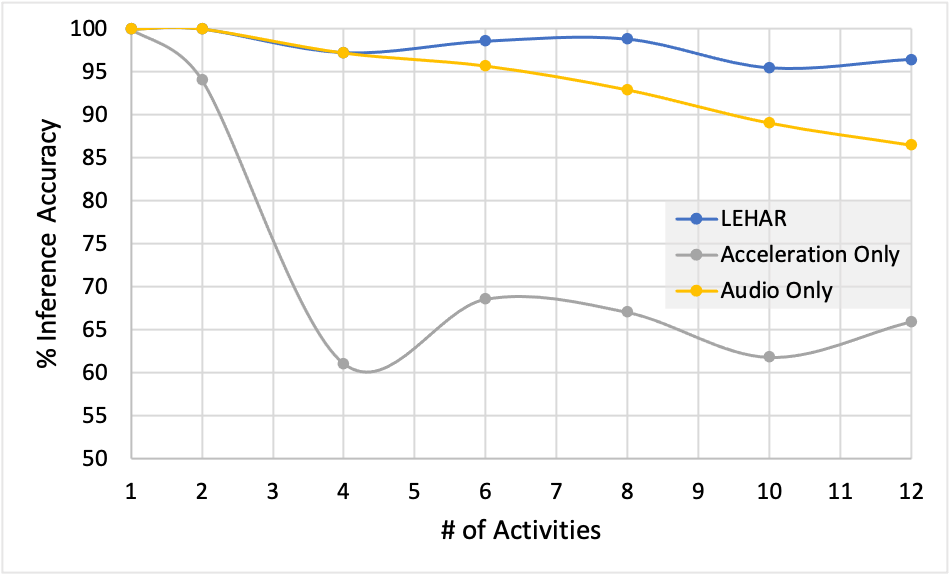}
   \caption{This figure shows how the accuracy of model varies with the number of activities.}
   \label{fig:scaling}
 \end{figure}
 
\subsection{Scaling with Activities}
\label{sec:scaling}

One key concern is whether a system can handle a large number of unique activities. To understand the scaling properties of \sys, I evaluated the accuracy of the system, adding one new activity at a time to the set of identifiable activities and retraining the model. I also trained an audio-only and acceleration-only model while varying the number of activities. The results of this experiment are shown in Figure~\ref{fig:scaling}. As the graph shows, the performance of the acceleration-only model degrades rapidly with the addition of activities and even the audio-only model degrades gradually. This is because the addition of activities introduces activities that have similar motions or sounds. In contrast, \sys's is able to provide consistently high accuracy despite the addition of more activities. I believe this is because the addition of location effectively limits the set of activities and associated sounds that the system is trying to match at any time to a small enough set for high accuracy.

 \begin{figure}[t]
   \centering
   \includegraphics[width=\linewidth]{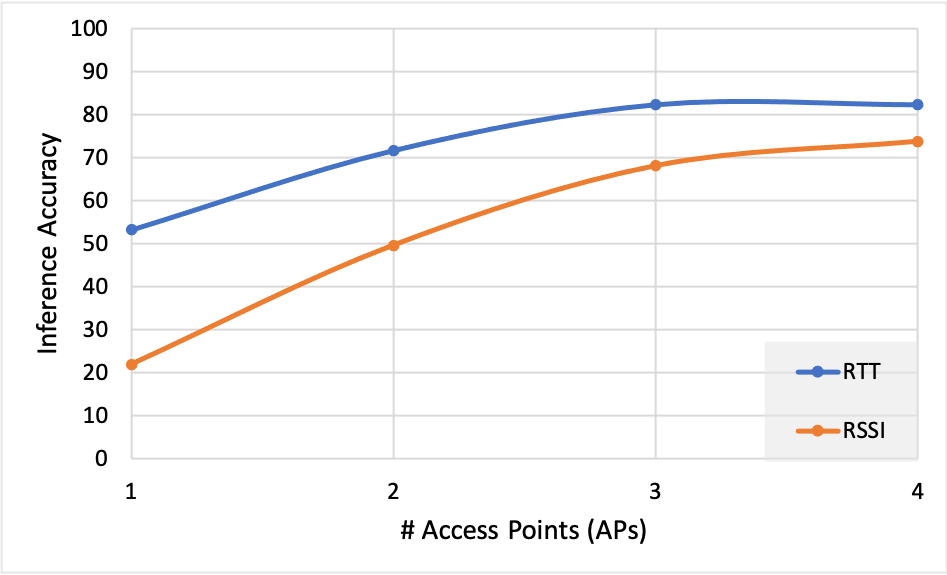}
   \caption{This figure shows how the accuracy of model varies with the number of access points and the use of RSSI instead of Wi-Fi RTT.}
   \label{fig:location}
 \end{figure}
 
\subsection{Impact of Location Accuracy}
\label{sec:locVacc}

In this section, I consider the impact of access point availability on accuracy of the whole system. In particular, I consider the issue of having fewer access points or having access points that don't support the Wi-Fi RTT protocol. To understand the value of additional access points, I varied the number of access points available in my training and testing data sets. I also considered the use of Received Signal Strength Indication (RSSI) instead of Wi-FI RTT information. As mentioned in Section~\ref{sec:rssi}, signal strength can be used for localization, but at some loss of accuracy. To clearly quantify the accuracy gained from the location information, I chose to train a location-only model for this experiment. Figure~\ref{fig:location} shows the results of this experiment. As the graph shows, the Wi-Fi RTT model performs significantly better than the RSSI-based model. This is likely because Wi-Fi RTT provides more accurate location information than RSSI. This gap decreases as you add more access points since the redundancy in information provided by multiple access points compensates for the inaccuracy of each measurement. In addition, it is worth noting that the accuracy of the Wi-Fi RTT system levels out at 3 access points. This is because, as mentioned in Section~\ref{sec:tdoa}, measurements to single access points localizes the user to a spherical shell, measurements to two access points localizes to a circular region, measurements to three access points localizes to two possible locations and to four access points localizes to a single spot. As a result, the gain in location information beyond 3 Wi-Fi RTT measurements is minimal. These measurements do indicate that even 2 Wi-Fi RTT access points or 3 standard access points for RSSI measurement provide significant value for HAR systems.

\section{Discussion}
\label{sec:discussion}

The results shared in the previous section indicate that \sys offers a viable method improving human activity recognition. Furthermore, the project proves that a easily deployable, accurate, and inexpensive solution can be created to address activity recognition inside homes. No prior instrumentation is necessary, allowing for an easy-to-implement solution. Although the project met the main constrains defined at the beginning of the paper, there are some limitations. These limitations include the following.

\begin{itemize}
\item \textbf{Activities.} Currently, \sys has been tested on twelve common activities. While these activities were clearly distinguishable, there may be activities that are more similar in audio. If the end goal is to help senior citizens, the system will have to be able to distinguish similar activities, such as taking medicines with water versus drinking orange juice. Further testing will be conducted as more activities are added.

\item \textbf{Hardware.} The \sys hardware was designed using technology that is not currently ubiquitous in homes. Wi-Fi RTT is still relatively new, and will most likely will be integrated into new products in the coming years.

\item \textbf{User testing.} Due to COVID-19, \sys was tested only by the author of the paper in the author's home. Future iterations will include having others conduct the activities to test the robustness of the system.
\end{itemize}


\section{Related Work}
\label{sec:related}

As mentioned earlier, Human Activity Recognition is not a new concept. Existing approaches fall into two broad categories: mobile device based and Smart Home based. At a very high level, the two approaches differ in the availability of activity recognition and the sensor information used. Mobile device-based designs are available as long as the user carries the devices, while smart home systems can only recognize activities when users are in the monitored spaces of the home. Smart Home designs benefit from the wide range sensors that can be deployed throughout the home, including cameras, microphones, motion detectors, etc. In contrast, mobile device designs only have access to the sensor information available in the mobile device. Therefore, smart home HAR systems result in much higher activity recognition accuracy than mobile devices-based systems. I describe some examples in each category below. 

\paragraph{Mobile Device HAR.} Limited forms of activity detection is deployed in a wide variety of mobile devices. For example, many smartphones keep track of walking and running. The Apple Watch~\cite{applewatch}, a widely deployed commercial device, automatically detects when you have begun specific types of  exercises~\cite{appleworkout}. This system only uses the accelerometer on the watch and is limited to detecting a small number of exercises including indoor walk, outdoor walk, indoor run, outdoor run, elliptical, rower, pool swim and open water swim. A similar system designed by D. Anguita, A. Ghio, et. al~\cite{2013.Anguita} used a dataset of accelerometer and gyroscope information from smartphones. Data was used to identify six activities that were all motion or posture oriented, such as standing, walking, and laying down. \sys improves upon both these designs by using audio and location data to perform accurate activity recognition on a larger variety of activities that are useful in a wider variety of applications. The ability to use additional sensor information also greatly improves the accuracy of \sys over the approaches used in these other systems. 

\begin{figure}[t]
  \centering
  \includegraphics[width=\linewidth]{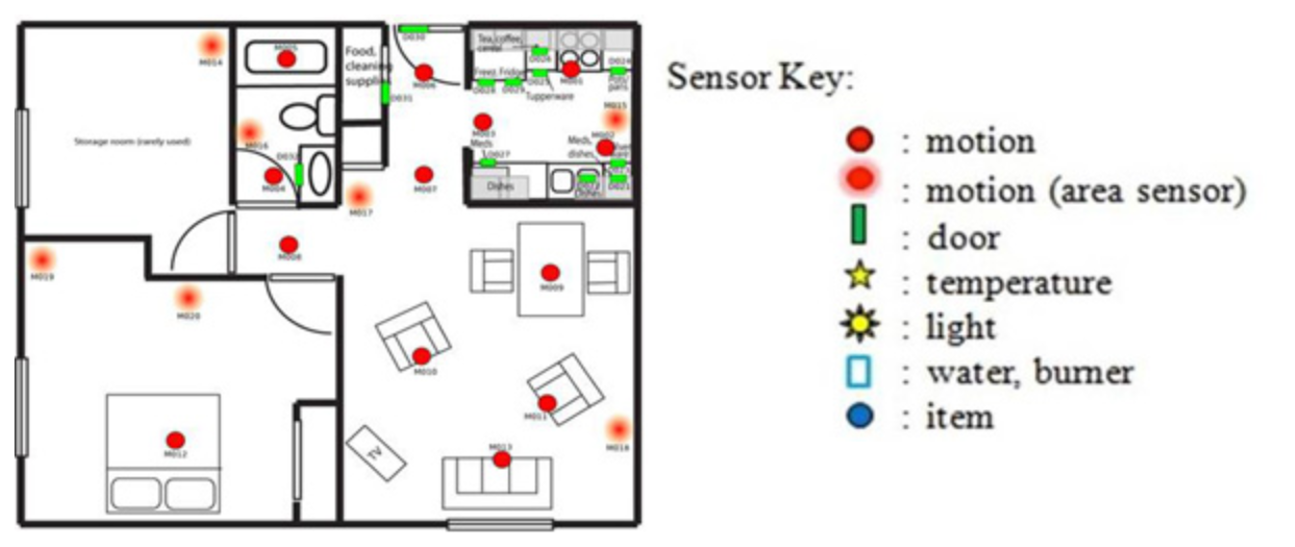}
  \caption{Layout of sensors in a Smart Home to enable activity recognition. Reproduced from~\cite{2014.Krishnan}.}
  \label{fig:smarthome}
\end{figure}

\paragraph{Smart Home HAR.} Most Smart Homes designs rely on external sensors placed in specific target objects: the microwave, the stove, the faucet, etc. This makes it relatively straightforward to recognize normal interactions with these devices such as turning on the faucet, cooking food, etc. These systems are typically limited to activities recognizable within the capabilities or coverage of these sensors. Furthermore, installing and maintaining the sensors in the locations you want to monitor adds to the overall cost of the system. Examples of such Smart Home projects include the Aware Home~\cite{aware}, CASAS~\cite{casas}, and PlaceLab~\cite{placelab}.  Figure~\ref{fig:smarthome} illustrates the sensor placement associated with a Smart Home HAR deployment. The figure demonstrates how sensors need to be actively placed for the system to work. The benefit of \sys is that it requires no extensive installation of sensors in the home.



\section{Conclusion}
\label{sec:conclusion}

The results of the study prove that high-accuracy human activity recognition is achievable using a combination of accelerometer, audio, and Wi-Fi Round Trip Time localization. This is a promising first step towards creating a simple, inexpensive, unobtrusive, and accurate system for monitoring human activity.

The \sys system fulfills the requirements determined in this project.

 \begin{itemize}
    \item The system is more accurate than existing systems at detecting human activity.
    \item No significant prior installation of materials or change to the room is required for activity monitoring.
    \item Extending the system to include more activities is easily automated by retraining the machine learning model.
 \end{itemize}

 Some of the next steps are to collect more data for each activity. The more data and the more variety, the more accurate the model will be.
 
 In addition, an important next step is to train new activities. The more activities the system can detect, the more useful it will be in the real world. The system has already proven that it can detect a larger variety of activities than current methods, but this can be taken further.

 Finally, other smartphone sensors can be implemented into the system to see how helpful they are to improving the accuracy.

%

\bibliographystyle{ACM-Reference-Format}
\bibliography{_references}


\begin{thebibliography}{24}


\ifx \showCODEN    \undefined \def \showCODEN     #1{\unskip}     \fi
\ifx \showDOI      \undefined \def \showDOI       #1{#1}\fi
\ifx \showISBNx    \undefined \def \showISBNx     #1{\unskip}     \fi
\ifx \showISBNxiii \undefined \def \showISBNxiii  #1{\unskip}     \fi
\ifx \showISSN     \undefined \def \showISSN      #1{\unskip}     \fi
\ifx \showLCCN     \undefined \def \showLCCN      #1{\unskip}     \fi
\ifx \shownote     \undefined \def \shownote      #1{#1}          \fi
\ifx \showarticletitle \undefined \def \showarticletitle #1{#1}   \fi
\ifx \showURL      \undefined \def \showURL       {\relax}        \fi
\providecommand\bibfield[2]{#2}
\providecommand\bibinfo[2]{#2}
\providecommand\natexlab[1]{#1}
\providecommand\showeprint[2][]{arXiv:#2}

\bibitem[\protect\citeauthoryear{{Ananda}, {Bernstein}, {Cunningham}, {Feess},
  and {Stroud}}{{Ananda} et~al\mbox{.}}{1990}]%
        {GPS}
\bibfield{author}{\bibinfo{person}{M.~P. {Ananda}}, \bibinfo{person}{H.
  {Bernstein}}, \bibinfo{person}{K.~E. {Cunningham}}, \bibinfo{person}{W.~A.
  {Feess}}, {and} \bibinfo{person}{E.~G. {Stroud}}.}
  \bibinfo{year}{1990}\natexlab{}.
\newblock \showarticletitle{Global Positioning System (GPS) autonomous
  navigation}. In \bibinfo{booktitle}{\emph{IEEE Symposium on Position Location
  and Navigation. A Decade of Excellence in the Navigation Sciences}}.
  \bibinfo{pages}{497--508}.
\newblock
\urldef\tempurl%
\url{https://doi.org/10.1109/PLANS.1990.66220}
\showDOI{\tempurl}


\bibitem[\protect\citeauthoryear{Anguita, Ghio, Oneto, Parra, and
  Reyes-Ortiz}{Anguita et~al\mbox{.}}{2013}]%
        {2013.Anguita}
\bibfield{author}{\bibinfo{person}{D. Anguita}, \bibinfo{person}{A. Ghio},
  \bibinfo{person}{L. Oneto}, \bibinfo{person}{X. Parra}, {and}
  \bibinfo{person}{J. Reyes-Ortiz}.} \bibinfo{year}{2013}\natexlab{}.
\newblock \showarticletitle{A Public Domain Dataset for Human Activity
  Recognition Using Smartphones}. In \bibinfo{booktitle}{\emph{Proceedings of
  ESANN}}. \bibinfo{pages}{437--442}.
\newblock
\urldef\tempurl%
\url{https://www.elen.ucl.ac.be/Proceedings/esann/esannpdf/es2013-84.pdf}
\showURL{%
\tempurl}


\bibitem[\protect\citeauthoryear{Apple}{Apple}{2021a}]%
        {appleworkout}
\bibfield{author}{\bibinfo{person}{Apple}.} \bibinfo{year}{2021}\natexlab{a}.
\newblock \bibinfo{title}{Use the Workout app on your Apple Watch}.
\newblock
\newblock
\urldef\tempurl%
\url{https://support.apple.com/en-us/HT204523#forgettostart}
\showURL{%
\tempurl}


\bibitem[\protect\citeauthoryear{Apple}{Apple}{2021b}]%
        {applewatch}
\bibfield{author}{\bibinfo{person}{Apple}.} \bibinfo{year}{2021}\natexlab{b}.
\newblock \bibinfo{title}{Watch - Apple}.
\newblock
\newblock
\urldef\tempurl%
\url{https://www.apple.com/watch/}
\showURL{%
\tempurl}


\bibitem[\protect\citeauthoryear{{Bahl} and {Padmanabhan}}{{Bahl} and
  {Padmanabhan}}{2000}]%
        {RADAR}
\bibfield{author}{\bibinfo{person}{P. {Bahl}} {and} \bibinfo{person}{V.~N.
  {Padmanabhan}}.} \bibinfo{year}{2000}\natexlab{}.
\newblock \showarticletitle{RADAR: an in-building RF-based user location and
  tracking system}. In \bibinfo{booktitle}{\emph{Proceedings IEEE INFOCOM 2000.
  Conference on Computer Communications. Nineteenth Annual Joint Conference of
  the IEEE Computer and Communications Societies (Cat. No.00CH37064)}},
  Vol.~\bibinfo{volume}{2}. \bibinfo{pages}{775--784 vol.2}.
\newblock
\urldef\tempurl%
\url{https://doi.org/10.1109/INFCOM.2000.832252}
\showDOI{\tempurl}


\bibitem[\protect\citeauthoryear{Bishop}{Bishop}{2006}]%
        {generalML}
\bibfield{author}{\bibinfo{person}{Christopher~M Bishop}.}
  \bibinfo{year}{2006}\natexlab{}.
\newblock \bibinfo{booktitle}{\emph{Pattern recognition and machine learning}}.
\newblock \bibinfo{publisher}{springer}.
\newblock


\bibitem[\protect\citeauthoryear{Bracewell and Bracewell}{Bracewell and
  Bracewell}{1986}]%
        {fourier}
\bibfield{author}{\bibinfo{person}{Ronald~Newbold Bracewell} {and}
  \bibinfo{person}{Ronald~N Bracewell}.} \bibinfo{year}{1986}\natexlab{}.
\newblock \bibinfo{booktitle}{\emph{The Fourier transform and its
  applications}}. Vol.~\bibinfo{volume}{31999}.
\newblock \bibinfo{publisher}{McGraw-Hill New York}.
\newblock


\bibitem[\protect\citeauthoryear{Cook, Schmitter-Edgecombe, Crandall, Sanders,
  and Thomas}{Cook et~al\mbox{.}}{2009}]%
        {casas}
\bibfield{author}{\bibinfo{person}{Diane Cook}, \bibinfo{person}{Maureen
  Schmitter-Edgecombe}, \bibinfo{person}{Aaron Crandall}, \bibinfo{person}{Chad
  Sanders}, {and} \bibinfo{person}{Brian Thomas}.}
  \bibinfo{year}{2009}\natexlab{}.
\newblock \showarticletitle{Collecting and disseminating smart home sensor data
  in the CASAS project}. In \bibinfo{booktitle}{\emph{Proceedings of the CHI
  workshop on developing shared home behavior datasets to advance HCI and
  ubiquitous computing research}}. \bibinfo{pages}{1--7}.
\newblock


\bibitem[\protect\citeauthoryear{{Gentner}, {Ulmschneider}, {Kuehner}, and
  {Dammann}}{{Gentner} et~al\mbox{.}}{2020}]%
        {wifirtt_accuracy}
\bibfield{author}{\bibinfo{person}{C. {Gentner}}, \bibinfo{person}{M.
  {Ulmschneider}}, \bibinfo{person}{I. {Kuehner}}, {and} \bibinfo{person}{A.
  {Dammann}}.} \bibinfo{year}{2020}\natexlab{}.
\newblock \showarticletitle{WiFi-RTT Indoor Positioning}. In
  \bibinfo{booktitle}{\emph{2020 IEEE/ION Position, Location and Navigation
  Symposium (PLANS)}}. \bibinfo{pages}{1029--1035}.
\newblock
\showISSN{2153-3598}
\urldef\tempurl%
\url{https://doi.org/10.1109/PLANS46316.2020.9110232}
\showDOI{\tempurl}


\bibitem[\protect\citeauthoryear{Google}{Google}{2021a}]%
        {nestwifi}
\bibfield{author}{\bibinfo{person}{Google}.} \bibinfo{year}{2021}\natexlab{a}.
\newblock \bibinfo{title}{Nest Wifi}.
\newblock
\newblock
\urldef\tempurl%
\url{https://store.google.com/product/nest_wifi}
\showURL{%
\tempurl}


\bibitem[\protect\citeauthoryear{Google}{Google}{2021b}]%
        {pixel4a}
\bibfield{author}{\bibinfo{person}{Google}.} \bibinfo{year}{2021}\natexlab{b}.
\newblock \bibinfo{title}{Pixel 4a Phones}.
\newblock
\newblock
\urldef\tempurl%
\url{https://store.google.com/product/pixel_4a}
\showURL{%
\tempurl}


\bibitem[\protect\citeauthoryear{Gouyon, Pachet, and Delerue}{Gouyon
  et~al\mbox{.}}{2000}]%
        {zerocrossing}
\bibfield{author}{\bibinfo{person}{Fabien Gouyon}, \bibinfo{person}{Francois
  Pachet}, {and} \bibinfo{person}{Olivier Delerue}.}
  \bibinfo{year}{2000}\natexlab{}.
\newblock \showarticletitle{On the Use of Zero-Crossing Rate for an Application
  of Classification of Percussive Sounds}. In
  \bibinfo{booktitle}{\emph{Proceedings of the COST G-6 Conference on Digital
  Audio Effects (DAFX-00)}}.
\newblock


\bibitem[\protect\citeauthoryear{Guide}{Guide}{2021}]%
        {androiddev}
\bibfield{author}{\bibinfo{person}{Android~Developer Guide}.}
  \bibinfo{year}{2021}\natexlab{}.
\newblock \bibinfo{title}{Wi-Fi location: ranging with RTT}.
\newblock
\newblock
\urldef\tempurl%
\url{https://developer.android.com/guide/topics/connectivity/wifi-rtt}
\showURL{%
\tempurl}


\bibitem[\protect\citeauthoryear{Intille, Larson, Tapia, Beaudin, Kaushik,
  Nawyn, and Rockinson}{Intille et~al\mbox{.}}{2006}]%
        {placelab}
\bibfield{author}{\bibinfo{person}{Stephen~S Intille}, \bibinfo{person}{Kent
  Larson}, \bibinfo{person}{Emmanuel~Munguia Tapia},
  \bibinfo{person}{Jennifer~S Beaudin}, \bibinfo{person}{Pallavi Kaushik},
  \bibinfo{person}{Jason Nawyn}, {and} \bibinfo{person}{Randy Rockinson}.}
  \bibinfo{year}{2006}\natexlab{}.
\newblock \showarticletitle{Using a live-in laboratory for ubiquitous computing
  research}. In \bibinfo{booktitle}{\emph{International Conference on Pervasive
  Computing}}. Springer, \bibinfo{pages}{349--365}.
\newblock


\bibitem[\protect\citeauthoryear{Kidd, Orr, Abowd, Atkeson, Essa, MacIntyre,
  Mynatt, Starner, and Newstetter}{Kidd et~al\mbox{.}}{1999}]%
        {aware}
\bibfield{author}{\bibinfo{person}{Cory~D Kidd}, \bibinfo{person}{Robert Orr},
  \bibinfo{person}{Gregory~D Abowd}, \bibinfo{person}{Christopher~G Atkeson},
  \bibinfo{person}{Irfan~A Essa}, \bibinfo{person}{Blair MacIntyre},
  \bibinfo{person}{Elizabeth Mynatt}, \bibinfo{person}{Thad~E Starner}, {and}
  \bibinfo{person}{Wendy Newstetter}.} \bibinfo{year}{1999}\natexlab{}.
\newblock \showarticletitle{The aware home: A living laboratory for ubiquitous
  computing research}. In \bibinfo{booktitle}{\emph{International workshop on
  cooperative buildings}}. Springer, \bibinfo{pages}{191--198}.
\newblock


\bibitem[\protect\citeauthoryear{Krishnan and Cook}{Krishnan and Cook}{2014}]%
        {2014.Krishnan}
\bibfield{author}{\bibinfo{person}{N Krishnan} {and} \bibinfo{person}{D.
  Cook}.} \bibinfo{year}{2014}\natexlab{}.
\newblock \showarticletitle{Activity Recognition on Streaming Sensor Data}. In
  \bibinfo{booktitle}{\emph{Proceedings of the 2014 Pervasive and Mobile
  Computing}}. \bibinfo{pages}{138–154}.
\newblock
\urldef\tempurl%
\url{https://www.ncbi.nlm.nih.gov/pmc/articles/PMC3979570/}
\showURL{%
\tempurl}


\bibitem[\protect\citeauthoryear{Ku{\l}akowski, Vales-Alonso, Egea-L{\'o}pez,
  Ludwin, and Garc{\'\i}a-Haro}{Ku{\l}akowski et~al\mbox{.}}{2010}]%
        {aoa1}
\bibfield{author}{\bibinfo{person}{Pawe{\l} Ku{\l}akowski},
  \bibinfo{person}{Javier Vales-Alonso}, \bibinfo{person}{Esteban
  Egea-L{\'o}pez}, \bibinfo{person}{Wies{\l}aw Ludwin}, {and}
  \bibinfo{person}{Joan Garc{\'\i}a-Haro}.} \bibinfo{year}{2010}\natexlab{}.
\newblock \showarticletitle{Angle-of-arrival localization based on antenna
  arrays for wireless sensor networks}.
\newblock \bibinfo{journal}{\emph{Computers \& Electrical Engineering}}
  \bibinfo{volume}{36}, \bibinfo{number}{6} (\bibinfo{year}{2010}),
  \bibinfo{pages}{1181--1186}.
\newblock


\bibitem[\protect\citeauthoryear{Nair and Hinton}{Nair and Hinton}{2010}]%
        {relu}
\bibfield{author}{\bibinfo{person}{Vinod Nair} {and}
  \bibinfo{person}{Geoffrey~E. Hinton}.} \bibinfo{year}{2010}\natexlab{}.
\newblock \showarticletitle{Rectified Linear Units Improve Restricted Boltzmann
  Machines}. In \bibinfo{booktitle}{\emph{Proceedings of the 27th International
  Conference on International Conference on Machine Learning}}
  \emph{(\bibinfo{series}{ICML'10})}. \bibinfo{publisher}{Omnipress},
  \bibinfo{address}{Madison, WI, USA}, \bibinfo{pages}{807–814}.
\newblock
\showISBNx{9781605589077}


\bibitem[\protect\citeauthoryear{of~IEEE Std 802.11-2012)}{of~IEEE Std
  802.11-2012)}{2016}]%
        {802.11mc}
\bibfield{author}{\bibinfo{person}{IEEE Std 802.11-2016~(Revision of IEEE Std
  802.11-2012)}.} \bibinfo{year}{2016}\natexlab{}.
\newblock \bibinfo{title}{IEEE Standard for Information technology
  Telecommunications and information exchange between systems Local and
  metropolitan area networks Specific requirements - Part 11: Wireless LAN
  Medium Access Control (MAC) and Physical Layer (PHY) Specifications}.
\newblock , \bibinfo{numpages}{3534}~pages.
\newblock


\bibitem[\protect\citeauthoryear{Peng and Sichitiu}{Peng and Sichitiu}{2006}]%
        {aoa2}
\bibfield{author}{\bibinfo{person}{Rong Peng} {and} \bibinfo{person}{Mihail~L
  Sichitiu}.} \bibinfo{year}{2006}\natexlab{}.
\newblock \showarticletitle{Angle of arrival localization for wireless sensor
  networks}. In \bibinfo{booktitle}{\emph{2006 3rd annual IEEE communications
  society on sensor and ad hoc communications and networks}},
  Vol.~\bibinfo{volume}{1}. Ieee, \bibinfo{pages}{374--382}.
\newblock


\bibitem[\protect\citeauthoryear{Reference}{Reference}{2021}]%
        {kerasapi}
\bibfield{author}{\bibinfo{person}{Keras~API Reference}.}
  \bibinfo{year}{2021}\natexlab{}.
\newblock \bibinfo{title}{Layer activation functions}.
\newblock
\newblock
\urldef\tempurl%
\url{https://keras.io/api/layers/activations/}
\showURL{%
\tempurl}


\bibitem[\protect\citeauthoryear{Sejdić, Djurović, and Jiang}{Sejdić
  et~al\mbox{.}}{2009}]%
        {STFT}
\bibfield{author}{\bibinfo{person}{Ervin Sejdić}, \bibinfo{person}{Igor
  Djurović}, {and} \bibinfo{person}{Jin Jiang}.}
  \bibinfo{year}{2009}\natexlab{}.
\newblock \showarticletitle{Time–frequency feature representation using
  energy concentration: An overview of recent advances}.
\newblock \bibinfo{journal}{\emph{Digital Signal Processing}}
  \bibinfo{volume}{19}, \bibinfo{number}{1} (\bibinfo{year}{2009}),
  \bibinfo{pages}{153--183}.
\newblock
\showISSN{1051-2004}
\urldef\tempurl%
\url{https://doi.org/10.1016/j.dsp.2007.12.004}
\showDOI{\tempurl}


\bibitem[\protect\citeauthoryear{Williams and Last}{Williams and Last}{2003}]%
        {williams2003loran}
\bibfield{author}{\bibinfo{person}{Paul Williams} {and} \bibinfo{person}{David
  Last}.} \bibinfo{year}{2003}\natexlab{}.
\newblock \showarticletitle{On Loran-C time-difference to Co-ordinate
  converters}. In \bibinfo{booktitle}{\emph{Proceedings-International Loran
  Association (ILA)-32nd Annual Convention and Technical Symposium}}.
  \bibinfo{pages}{3--7}.
\newblock


\bibitem[\protect\citeauthoryear{Xu, Duan, Cai, Chia, Xu, and Tian}{Xu
  et~al\mbox{.}}{2004}]%
        {MFCC}
\bibfield{author}{\bibinfo{person}{Min Xu}, \bibinfo{person}{Ling-Yu Duan},
  \bibinfo{person}{Jianfei Cai}, \bibinfo{person}{Liang-Tien Chia},
  \bibinfo{person}{Changsheng Xu}, {and} \bibinfo{person}{Qi Tian}.}
  \bibinfo{year}{2004}\natexlab{}.
\newblock \showarticletitle{HMM-based audio keyword generation}. In
  \bibinfo{booktitle}{\emph{Pacific-Rim Conference on Multimedia}}. Springer,
  \bibinfo{pages}{566--574}.
\newblock


\end{thebibliography}

\newpage
\onecolumn
\appendix

\section{Code}
\label{sec:code}
\subsection{Data Collection}
\label{sec:collectionCode}

\begin{lstlisting}[language=Java]
package com.example.readsensors;

import androidx.annotation.RequiresApi;
import androidx.appcompat.app.AppCompatActivity;
import androidx.core.app.ActivityCompat;
import android.content.BroadcastReceiver;
import android.content.ContentResolver;
import android.content.ContentValues;
import android.content.Context;
import android.content.Intent;
import android.content.IntentFilter;
import android.content.pm.PackageManager;
import android.hardware.SensorEventListener;
import android.hardware.SensorManager;
import android.media.MediaRecorder;
import android.net.Uri;
import android.net.wifi.ScanResult;
import android.net.wifi.WifiInfo;
import android.net.wifi.WifiManager;
import android.net.wifi.rtt.RangingRequest;
import android.net.wifi.rtt.RangingResult;
import android.net.wifi.rtt.RangingResultCallback;
import android.net.wifi.rtt.WifiRttManager;
import android.os.Build;
import android.os.Bundle;
import android.hardware.Sensor;
import android.hardware.SensorEvent;
import android.os.Environment;
import android.provider.MediaStore;
import android.util.Log;
import android.view.View;
import android.widget.ArrayAdapter;
import android.widget.TextView;
import android.widget.Toast;
import java.io.File;
import java.io.FileNotFoundException;
import java.io.FileOutputStream;
import java.io.IOException;
import java.util.ArrayList;
import java.util.Hashtable;
import java.util.List;
import java.util.concurrent.Executor;

public class MainActivity extends AppCompatActivity implements SensorEventListener {
    private SensorManager senSensorManager;
    private Sensor senAccelerometer;
    public static String data = "X, Y, Z, Basement, Kitchen, Upstairs, Dining Room" + "\r\n";
    ArrayList<Integer> xArray = new ArrayList<Integer>();
    ArrayList<Integer> yArray = new ArrayList<Integer>();
    ArrayList<Integer> zArray = new ArrayList<Integer>();
    public static float x;
    public static float y;
    public static float z;
    public static String rssi;
    public static boolean record = false;
    public static String D1;
    public static String D2;
    public static String D3;
    public static String D4;
    public static int save = 0;
    public static File audiofile;
    public static long current;
    long start;

    @Override
    protected void onCreate(Bundle savedInstanceState) {
        super.onCreate(savedInstanceState);
        setContentView(R.layout.activity_main);

        senSensorManager = (SensorManager) getSystemService(Context.SENSOR_SERVICE);
        senAccelerometer = senSensorManager.getDefaultSensor(Sensor.TYPE_ACCELEROMETER);
        senSensorManager.registerListener(this, senSensorManager.getDefaultSensor(Sensor.TYPE_ACCELEROMETER), SensorManager.SENSOR_DELAY_NORMAL);
    }

    Context context;
    MediaRecorder recorder = new MediaRecorder();

    @Override
    public void onSensorChanged(SensorEvent event) {
        Sensor mySensor = event.sensor;
        if (mySensor.getType() == Sensor.TYPE_ACCELEROMETER) {
            x = event.values[0];
            y = event.values[1];
            z = event.values[2];
            TextView textViewx = findViewById(R.id.textViewX);
            textViewx.setText(new String(String.valueOf(x)));
            TextView textViewy = findViewById(R.id.textViewY);
            textViewy.setText(new String(String.valueOf(y)));
            TextView textViewz = findViewById(R.id.textViewZ);
            textViewz.setText(new String(String.valueOf(z)));
        }
        context = getApplicationContext();
        wifiManager = (WifiManager) context.getSystemService(Context.WIFI_SERVICE);
        wifiScan(null);

        if ((save == 0) && (recorder != null) && record) {
            File dir = Environment.getExternalStoragePublicDirectory(Environment.DIRECTORY_DOWNLOADS);
            try {
                audiofile = File.createTempFile("sound" +(current/1000), ".wav", dir);
            } catch (IOException e) {
                System.out.println("audio not working");
            }

            recorder.setAudioSource(MediaRecorder.AudioSource.MIC);
            recorder.setOutputFormat(MediaRecorder.OutputFormat.THREE_GPP);
            recorder.setAudioEncoder(MediaRecorder.AudioEncoder.AMR_NB);
            recorder.setOutputFile(audiofile.getAbsolutePath());
            try {
                recorder.prepare();
            } catch (IOException e) {
                e.printStackTrace();
            }

            recorder.start();

            start = System.currentTimeMillis();
            save = 1;
        }

        if ((System.currentTimeMillis() >= start + 3000) && (save != 0) && record) {
            recorder.stop();
            addRecordingToMediaLibrary();
            save = 0;
            int total = 0;
            int Xavg = 0;
            for(int i = 0; i < xArray.size(); i++)
            {
                total += Math.abs(xArray.get(i));
                Xavg = total / xArray.size();
            }
            total = 0;
            int Yavg = 0;
            for(int i = 0; i < yArray.size(); i++)
            {
                total += Math.abs(yArray.get(i));
                Yavg = total / yArray.size();
            }
            total = 0;
            int Zavg = 0;
            for(int i = 0; i < zArray.size(); i++)
            {
                total += Math.abs(zArray.get(i));
                Zavg = total / zArray.size();
            }

            data = data + new String(String.valueOf(Xavg)) + "," + new String(String.valueOf(Yavg)) + "," + new String(String.valueOf(Zavg)) + "," + D1 + "," + D2 + "," + D3 + "," + D4 +"\n";

            xArray.clear();
            yArray.clear();
            zArray.clear();
        }

        if (record) {
            xArray.add((int) x);
            yArray.add((int) y);
            zArray.add((int) z);

        }
    }

    protected void addRecordingToMediaLibrary() {
        ContentValues values = new ContentValues(4);
        current = System.currentTimeMillis();
        values.put(MediaStore.Audio.Media.TITLE, "audio" + audiofile.getName());
        values.put(MediaStore.Audio.Media.DATE_ADDED, (int) (current / 1000));
        values.put(MediaStore.Audio.Media.MIME_TYPE, "audio/3gpp");
        values.put(MediaStore.Audio.Media.DATA, audiofile.getAbsolutePath());
        ContentResolver contentResolver = getContentResolver();
        Uri base = MediaStore.Audio.Media.EXTERNAL_CONTENT_URI;
        Uri newUri = contentResolver.insert(base, values);
        sendBroadcast(new Intent(Intent.ACTION_MEDIA_SCANNER_SCAN_FILE, newUri));
        Toast.makeText(this, "Added File " + newUri, Toast.LENGTH_LONG).show();
    }

    @Override
    public void onAccuracyChanged(Sensor sensor, int accuracy) {
    }

    protected void onPause() {
        super.onPause();
    }

    protected void onResume() {
        super.onResume();
    }

    public void startRecording(View view) {
        record = true;
    }

    public void stopRecording(View view) {
        record = false;
        data = data + "\r\n";
    }

    public void sendData(View view) throws IOException {
        String FILENAME = "phone_data" + System.currentTimeMillis() / 1000L + ".csv";

        File folder = Environment.getExternalStoragePublicDirectory(Environment.DIRECTORY_DOWNLOADS);
        File myFile = new File(folder, FILENAME);
        FileOutputStream fstream = new FileOutputStream(myFile);
        fstream.write(data.getBytes());
        fstream.close();
        TextView textViewMessage = findViewById(R.id.textViewMessage);
        textViewMessage.setText("worked");
    }

    private List<ScanResult> results;
    private ArrayList<String> arrayList = new ArrayList<>();
    private ArrayAdapter adapter;
    public WifiManager wifiManager;

    public void wifiScan(View view) {

        IntentFilter intentFilter = new IntentFilter();
        intentFilter.addAction(WifiManager.SCAN_RESULTS_AVAILABLE_ACTION);
        context.registerReceiver(wifiScanReceiver, intentFilter);

        boolean success = wifiManager.startScan();
        if (!success) {
            scanFailure();
        }
    }

    BroadcastReceiver wifiScanReceiver = new BroadcastReceiver() {
        @RequiresApi(api = Build.VERSION_CODES.P)
        @Override
        public void onReceive(Context c, Intent intent) {
            boolean success = intent.getBooleanExtra(
                    WifiManager.EXTRA_RESULTS_UPDATED, false);
            if (success) {
                scanSuccess();
            } else {
                scanFailure();
            }
        }
    };
    Hashtable<Integer, String> levels = new Hashtable<Integer, String>();

    @RequiresApi(api = Build.VERSION_CODES.P)
    private void scanSuccess() {
        System.out.println("Scan Success");
        List<ScanResult> results = wifiManager.getScanResults();
        ScanResult scanResult = null;
        ScanResult scanResult2 = null;
        ScanResult scanResult3 = null;
        ScanResult scanResult4 = null;
        int z = 0;
        for (int j = 0; j < results.size(); j++) {
            if (results.get(j).SSID.equals("HomeG") && (results.get(j).is80211mcResponder())) {
                levels.put(z, results.get(j).BSSID);
                if (z == 0) {
                    scanResult = results.get(j);
                }
                if (z == 1) {
                    scanResult2 = results.get(j);
                }
                if (z == 2) {
                    scanResult3 = results.get(j);
                }
                if (z == 3) {
                    scanResult4 = results.get(j);
                }
                z++;
            }
        }
        WifiRttManager mgr = (WifiRttManager) context.getSystemService(Context.WIFI_RTT_RANGING_SERVICE);
        if (scanResult != null && scanResult2 != null && scanResult3 != null && scanResult4 != null) {
            final RangingRequest request = new RangingRequest.Builder()
                    .addAccessPoint(scanResult)
                    .addAccessPoint(scanResult2)
                    .addAccessPoint(scanResult3)
                    .addAccessPoint(scanResult4)
                    .build();
            final RangingResultCallback callback = new RangingResultCallback() {
                public void onRangingResults(List<RangingResult> resultsRTT) {
                    System.out.println(resultsRTT);
                    System.out.println("Ranging Result");
                    try {
                        for (int k = 0; k < 4; k++) {
                            if (String.valueOf(resultsRTT.get(k).getMacAddress()).equals("b0:e4:d5:04:8a:c5")) {
                                System.out.println(resultsRTT.get(k).getMacAddress() + ": " + resultsRTT.get(k).getDistanceMm());
                                D1 = String.valueOf(resultsRTT.get(k).getDistanceMm());
                                TextView textViewD1 = findViewById(R.id.textViewD1);
                                textViewD1.setText(D1);
                                break;
                            }
                        }
                        for (int k = 0; k < 4; k++) {
                            if (String.valueOf(resultsRTT.get(k).getMacAddress()).equals("f0:72:ea:48:bc:95")) {
                                System.out.println(resultsRTT.get(k).getMacAddress() + ": " + resultsRTT.get(k).getDistanceMm());
                                D2 = String.valueOf(resultsRTT.get(k).getDistanceMm());
                                TextView textViewD2 = findViewById(R.id.textViewD2);
                                textViewD2.setText(D2);
                                break;
                            }
                        }
                        for (int k = 0; k < 4; k++) {
                            if (String.valueOf(resultsRTT.get(k).getMacAddress()).equals("cc:f4:11:4a:49:c4")) {
                                System.out.println(resultsRTT.get(k).getMacAddress() + ": " + resultsRTT.get(k).getDistanceMm());
                                D3 = String.valueOf(resultsRTT.get(k).getDistanceMm());
                                TextView textViewD3 = findViewById(R.id.textViewD3);
                                textViewD3.setText(D3);
                                break;
                            }
                        }
                        for (int k = 0; k < 4; k++) {
                            if (String.valueOf(resultsRTT.get(k).getMacAddress()).equals("b0:e4:d5:17:63:65")) {
                                System.out.println(resultsRTT.get(k).getMacAddress() + ": " + resultsRTT.get(k).getDistanceMm());
                                D4 = String.valueOf(resultsRTT.get(k).getDistanceMm());
                                TextView textViewD4 = findViewById(R.id.textViewD4);
                                textViewD4.setText(D4);
                                break;
                            }
                        }
                    } catch(Exception e) {
                        System.out.println(e);
                        TextView textViewD1 = findViewById(R.id.textViewD1);
                    }

                }

                public void onRangingFailure(int code) {
                    // Handle failure
                    List<ScanResult> results = wifiManager.getScanResults();
                }
            };
            if (ActivityCompat.checkSelfPermission(this, android.Manifest.permission.ACCESS_FINE_LOCATION) != PackageManager.PERMISSION_GRANTED) {
                return;
            }
            final Executor mainExecutor;
            mainExecutor = context.getMainExecutor();
            mgr.startRanging(request, mainExecutor, callback);
        }
    }

    private void scanFailure() {
        System.out.println("not working");
        List<ScanResult> results = wifiManager.getScanResults();
    }

}
\end{lstlisting}

\subsection{Feature Extraction}
\label{sec:extractioncode}

\begin{lstlisting}[language=Python]
import librosa
import pandas as pd
import numpy as np
import matplotlib.pyplot as plt
%matplotlib inline
import os
from PIL import Image
import pathlib
import csv
import librosa
import librosa.display
from sklearn.model_selection import train_test_split
from sklearn.preprocessing import LabelEncoder, StandardScaler
import keras
import warnings
warnings.filterwarnings('ignore')

all_files = ['PhoneAudio/phone_datatyping.csv', 'PhoneAudio/phone_datavaccuming.csv',
             'PhoneAudio/phone_datavaccuming2.csv', 'PhoneAudio/phone_datafaucet.csv'
             'PhoneAudio/phone_datafaucet2.csv', 'PhoneAudio/phone_databrushingTeeth.csv',
             'PhoneAudio/phone_databrushingTeeth2.csv', 'PhoneAudio/phone_datacoffee.csv',
             'PhoneAudio/phone_datacoffee2.csv', 'PhoneAudio/phone_datamedicine.csv',
             'PhoneAudio/phone_datashaving.csv', 'PhoneAudio/phone_datacooking1.csv',
             'PhoneAudio/phone_datacooking2.csv', 'PhoneAudio/phone_datafalse.csv']
             
df_from_each_file = (pd.read_csv(f, sep=',') for f in all_files)
df_merged   = pd.concat(df_from_each_file, ignore_index=True)
df_merged.to_csv( "phone_data.csv")

header = 'filename chroma_stft rmse spectral_centroid spectral_bandwidth rolloff zero_crossing_rate'
for i in range(1, 21):
    header += f' mfcc{i}'
header += ' label'
header = header.split()

file = open('data.csv', 'w', newline='')
with file:
    writer = csv.writer(file)
    writer.writerow(header)
    
activities = 'typing vaccuming faucet brushingTeeth coffee medicine shaving cooking false'.split()

for a in activities:
    for filename in os.listdir(f'./PhoneAudio/{a}'):
        audioname = f'./PhoneAudio/{a}/{filename}'
        y, sr = librosa.load(audioname, mono=True, duration=3)
        chroma_stft = librosa.feature.chroma_stft(y=y, sr=sr)
        rmse = librosa.feature.rmse(y=y)[0]
        spec_cent = librosa.feature.spectral_centroid(y=y, sr=sr)
        spec_bw = librosa.feature.spectral_bandwidth(y=y, sr=sr)
        rolloff = librosa.feature.spectral_rolloff(y=y, sr=sr)
        zcr = librosa.feature.zero_crossing_rate(y)
        mfcc = librosa.feature.mfcc(y=y, sr=sr)
        to_append = f'{filename} {np.mean(chroma_stft)} {np.mean(rmse)} {np.mean(spec_cent)} {np.mean(spec_bw)} {np.mean(rolloff)} {np.mean(zcr)}'    
        for e in mfcc:
            to_append += f' {np.mean(e)}'
        to_append += f' {a}'
        file = open('data.csv', 'a', newline='')
        with file:
            writer = csv.writer(file)
            writer.writerow(to_append.split())

data = pd.read_csv('data.csv')

# Dropping unused columns and merging data
data = data.drop(['filename'],axis=1)
genre_list = data.iloc[:, -1]
loc = pd.read_csv("phone_data.csv")
merged = pd.concat([data, loc], axis=1)
merged.to_csv("data.csv", index=False)
data = pd.read_csv('data.csv')


encoder = LabelEncoder()
y = encoder.fit_transform(genre_list)

data=data.drop(['label'],axis=1)
data=data.drop(['Unnamed: 0'],axis=1)
scaler = StandardScaler()
scaler.fit(np.array(data.iloc[:]))
X = scaler.transform(np.array(data.iloc[:]))

\end{lstlisting}

\subsection{Model Training}
\label{sec:trainingcode}

\begin{lstlisting}[language=Python]
X_train, X_test, y_train, y_test = train_test_split(X, y, test_size=0.2, random_state=1)

from keras import models
from keras import layers

model = models.Sequential()
model.add(layers.Dense(256, activation='relu', input_shape=(X_train.shape[1],)))
model.add(layers.Dense(128, activation='relu'))
model.add(layers.Dense(64, activation='relu'))
model.add(layers.Dense(32, activation='relu'))
model.add(layers.Dense(16, activation='softmax'))
model.add(layers.Dense(9, activation='softmax'))
model.compile(optimizer='adam',
              loss='sparse_categorical_crossentropy',
              metrics=['accuracy'])
history = model.fit(X_train,
                    y_train,
                    epochs=1000,
                    validation_data=(X_test, y_test))
                    
\end{lstlisting}

\subsection{Activity Inference}
\label{sec:inferencecode}

\begin{lstlisting}[language=Python]
header = 'filename chroma_stft rmse spectral_centroid spectral_bandwidth rolloff zero_crossing_rate'
for i in range(1, 21):
    header += f' mfcc{i}'
header = header.split()

file = open('testdata.csv', 'w', newline='')
with file:
    writer = csv.writer(file)
    writer.writerow(header)

for filename in os.listdir(f'./testData/typing'):
    audioname = f'./testData/typing/{filename}'
    y, sr = librosa.load(audioname, mono=True, duration=3)
    chroma_stft = librosa.feature.chroma_stft(y=y, sr=sr)
    rmse = librosa.feature.rmse(y=y)[0]
    spec_cent = librosa.feature.spectral_centroid(y=y, sr=sr)
    spec_bw = librosa.feature.spectral_bandwidth(y=y, sr=sr)
    rolloff = librosa.feature.spectral_rolloff(y=y, sr=sr)
    zcr = librosa.feature.zero_crossing_rate(y)
    mfcc = librosa.feature.mfcc(y=y, sr=sr)
    to_append = f'{filename} {np.mean(chroma_stft)} {np.mean(rmse)} {np.mean(spec_cent)} {np.mean(spec_bw)} {np.mean(rolloff)} {np.mean(zcr)}'    
    for e in mfcc:
        to_append += f' {np.mean(e)}'
    file = open('testdata.csv', 'a', newline='')
    with file:
        writer = csv.writer(file)
        writer.writerow(to_append.split())
        
data = pd.read_csv('testdata.csv')

# Dropping unneccesary columns
data = data.drop(['filename'],axis=1)
loc = pd.read_csv("testData/testphone_datatyping.csv")
merged = pd.concat([data, loc], axis=1)
merged.to_csv("testdata.csv", index=False)
data = pd.read_csv('testdata.csv')

X = scaler.transform(np.array(data.iloc[:]))
results = model.predict(X)

for element in results:
    print(np.where(element == max(element)), np.where(element == np.unique(element)[-2]))

\end{lstlisting}

\end{document}